\documentstyle[rotate,emulateapj]{article}

\newcommand{\etal}{et~al.\ }
\newcommand{\CIVdblt}{{\rm C}\kern 0.1em{\sc iv}~$\lambda\lambda 1548, 1550$}
\newcommand{\MgIIdblt}{{\rm Mg}\kern 0.1em{\sc ii}~$\lambda\lambda 2796, 2803$}
\newcommand{\NVdblt}{{\rm N}\kern 0.1em{\sc v}~$\lambda\lambda 1238, 1242$}  
\newcommand{\SVIdblt}{{\rm S}\kern 0.1em{\sc vi}~$\lambda\lambda 933, 944$} 
\newcommand{\OVIdblt}{{\rm O}\kern 0.1em{\sc vi}~$\lambda\lambda 1031, 1037$} 
\newcommand{\SiIIdblt}{{\rm Si}\kern 0.1em{\sc ii}~$\lambda\lambda1190, 1193$}  
\newcommand{\SiIVdblt}{{\rm Si}\kern 0.1em{\sc iv}~$\lambda\lambda1393, 1402$} 
\newcommand{\AlI}{\hbox{{\rm Al}\kern 0.1em{\sc i}}}
\newcommand{\AlII}{\hbox{{\rm Al}\kern 0.1em{\sc ii}}}
\newcommand{\AlIII}{{\hbox{\rm Al}\kern 0.1em{\sc iii}}}
\newcommand{\CaII}{\hbox{{\rm Ca}\kern 0.1em{\sc ii}}}
\newcommand{\CII}{\hbox{{\rm C}\kern 0.1em{\sc ii}}}
\newcommand{\CIIe}{\hbox{{\rm C$^{\ast}$}\kern 0.1em{\sc ii}}}
\newcommand{\CIII}{\hbox{{\rm C}\kern 0.1em{\sc iii}}}
\newcommand{\CIV}{\hbox{{\rm C}\kern 0.1em{\sc iv}}}
\newcommand{\CV}{\hbox{{\rm C}\kern 0.1em{\sc v}}}
\newcommand{\HI}{\hbox{{\rm H}\kern 0.1em{\sc i}}}
\newcommand{\HII}{\hbox{{\rm H}\kern 0.1em{\sc ii}}}
\newcommand{\Lya}{\hbox{{\rm Ly}\kern 0.1em$\alpha$}}
\newcommand{\Lyb}{\hbox{{\rm Ly}\kern 0.1em$\beta$}}
\newcommand{\Lyg}{\hbox{{\rm Ly}\kern 0.1em$\gamma$}}
\newcommand{\Lyd}{\hbox{{\rm Ly}\kern 0.1em$\delta$}}
\newcommand{\Lye}{\hbox{{\rm Ly}\kern 0.1em$\epsilon$}}
\newcommand{\Lyphi}{\hbox{{\rm Ly}\kern 0.1em$\phi$}}
\newcommand{\Lyfive}{\hbox{{\rm Ly}\kern 0.1em$5$}}
\newcommand{\Lysix}{\hbox{{\rm Ly}\kern 0.1em$6$}}
\newcommand{\Lyseven}{\hbox{{\rm Ly}\kern 0.1em$7$}}
\newcommand{\Lyeight}{\hbox{{\rm Ly}\kern 0.1em$8$}}
\newcommand{\Lynine}{\hbox{{\rm Ly}\kern 0.1em$9$}}
\newcommand{\Lyten}{\hbox{{\rm Ly}\kern 0.1em$10$}}
\newcommand{\HeI}{\hbox{{\rm He}\kern 0.1em{\sc i}}}
\newcommand{\HeII}{\hbox{{\rm He}\kern 0.1em{\sc ii}}}
\newcommand{\FeI}{\hbox{{\rm Fe}\kern 0.1em{\sc i}}}
\newcommand{\FeII}{\hbox{{\rm Fe}\kern 0.1em{\sc ii}}}
\newcommand{\FeIII}{\hbox{{\rm Fe}\kern 0.1em{\sc iii}}}
\newcommand{\MnII}{\hbox{{\rm Mn}\kern 0.1em{\sc ii}}}
\newcommand{\MgI}{\hbox{{\rm Mg}\kern 0.1em{\sc i}}}
\newcommand{\MgII}{\hbox{{\rm Mg}\kern 0.1em{\sc ii}}}
\newcommand{\MgIII}{\hbox{{\rm Mg}\kern 0.1em{\sc iii}}}
\newcommand{\NI}{\hbox{{\rm N}\kern 0.1em{\sc i}}}
\newcommand{\NII}{\hbox{{\rm N}\kern 0.1em{\sc ii}}}
\newcommand{\NIII}{\hbox{{\rm N}\kern 0.1em{\sc iii}}}
\newcommand{\NV}{\hbox{{\rm N}\kern 0.1em{\sc v}}}
\newcommand{\OVI}{\hbox{{\rm O}\kern 0.1em{\sc vi}}}
\newcommand{\OI}{\hbox{{\rm O}\kern 0.1em{\sc i}}}
\newcommand{\OII}{\hbox{[{\rm O}\kern 0.1em{\sc ii}]}}
\newcommand{\SVI}{{\rm S}\kern 0.1em{\sc vi}}
\newcommand{\SiI}{\hbox{{\rm Si}\kern 0.1em{\sc i}}}
\newcommand{\SiII}{\hbox{{\rm Si}\kern 0.1em{\sc ii}}}
\newcommand{\SiIII}{\hbox{{\rm Si}\kern 0.1em{\sc iii}}}
\newcommand{\SiIV}{\hbox{{\rm Si}\kern 0.1em{\sc iv}}}
\newcommand{\SII}{\hbox{{\rm S}\kern 0.1em{\sc ii}}}
\newcommand{\SIII}{\hbox{{\rm S}\kern 0.1em{\sc iii}}}
\newcommand{\NaI}{\hbox{{\rm Na}\kern 0.1em{\sc i}}}
\newcommand{\kms}{\hbox{km~s$^{-1}$}}
\newcommand{\cmsq}{\hbox{cm$^{-2}$}}
\newcommand{\cc}{\hbox{cm$^{-3}$}}

\tighten
\submitted{The Astronomical Journal, {\it in press}}
\begin{document}


\lefthead{CHURCHILL \& CHARLTON}
\righthead{MULTIPLE PHASES OF HALO GAS}


\title{
The Multiple Phases of Interstellar and Halo Gas \\ in a
Possible Group of Galaxies at $\lowercase{z} \sim 1$\altaffilmark{1}}

\author{Christopher~W.~Churchill\altaffilmark{2}, 
        and 
        Jane~C.~Charlton\altaffilmark{3}}

\affil{Astronomy and Astrophysics Department \\
       Pennsylvania State University,
       University Park, PA 16802 \\
       {\it cwc, charlton@astro.psu.edu}}

\altaffiltext{1}{Based in part on observations obtained at the
W.~M. Keck Observatory, which is jointly operated by the University of
California and the California Institute of Technology. Based in part
on observations obtained with the NASA/ESA {\it Hubble Space
Telescope}, which is operated by the STScI for the Association of
Universities for Research in Astronomy, Inc., under NASA contract
NAS5--26555.}
\altaffiltext{2}{Visiting Astronomer, The W.~M.~Keck Observatory}
\altaffiltext{3}{Center for Gravitational Physics and Geometry}


\begin{abstract}

We used HIRES/Keck profiles ($R\sim 6$~{\kms}) of {\MgII} and {\FeII}
in combination with FOS/{\it HST\/} spectra  ($R\sim 230$~{\kms}) to
place constraints on the physical conditions (metallicities,
ionization conditions, and multiphase distribution) of absorbing gas
in three galaxies at $z=0.9254$, $0.9276$, and $0.9343$ along the line
of sight to PG~$1206+459$.
The chemical and ionization species covered in the FOS/{\it HST\/}
spectra are {\HI}, {\SiII}, {\CII}, {\NII}, {\FeIII}, {\SiIII},
{\SiIV}, {\NIII}, {\CIII}, {\CIV}, {\SVI}, {\NV}, and {\OVI}, with
ionization potentials ranging from 13.6 to 138 eV.
The multiple {\MgII} clouds exhibit complex kinematics and the {\CIV},
{\NV} and {\OVI} are exceptionally strong in absorption.
We assumed that the {\MgII} clouds are photoionized by the
extra--galactic background and determined the allowed ranges of their
physical properties as constrained by the absorption strengths in the
FOS spectra.
A main result of this paper is that the low resolution spectra can
provide meaningful {\it constraints\/} on the physical conditions of
the {\MgII} clouds, including allowed ranges of cloud to cloud
variations within a system.
We find that the {\MgII} clouds, which have a typical size of
$\sim 100$~pc, give rise to the {\SiIV}, the majority of which arises
in a single, very large ($\sim 5$~kpc), higher ionization cloud.
However, the {\MgII} clouds cannot account for the strong {\CIV},
{\NV}, and {\OVI} absorption.
We conclude that the {\MgII} clouds are embedded in extended
($10$--$20$~kpc), highly ionized gas that gives rise to {\CIV}, {\NV},
and {\OVI}; these are multiphase absorption systems.
The high ionization phases have near--solar metallicity and are
consistent with Galactic--like coronae surrounding the individual
galaxies, as opposed to a very extended common ``halo'' encompassing
all three galaxies.

\end{abstract}

\keywords{galaxies: structure --- galaxies: evolution
      --- galaxies: halos     --- galaxies: abundances
      --- quasars: absorption lines}

\section{Introduction}
\label{sec:intro}

An ultimate goal of the study of quasar (QSO) absorption lines is to
develop a comprehensive understanding of the kinematic,
chemical, and ionization conditions of gaseous structures in
early--epoch galaxies and to chart their cosmic evolution.
For a comprehensive physical picture of any given absorption
system, both high resolution spectra of a wide range of chemical and
ionization species and the empirically measured properties of the
associated galaxies are required.
For $z\sim 1$, shortly following the epoch of peak star formation, the
association between {\MgIIdblt} absorption and galaxies is well
established (\cite{bb91}; \cite{steidel95}), and their kinematics,
though complex and varied, are consistent with being coupled to the
galaxies themselves (\cite{csv96}; \cite{cc98}).
It is unfortunate, however, that for $z \sim 1$, the spectroscopic
data of are not of uniform, high quality due to the need for large
amounts of space--based telescope time to observe ultraviolet
wavelengths.
Presently, any comprehensive analyses of low redshift systems for
which the low ionization species (i.e.\ {\MgII}, {\FeII}, {\MgI}) have
been observed at high resolution with HIRES/Keck (see Churchill, Vogt,
\& Charlton 1999b\nocite{cvc99}) must incorporate low resolution
FOS/{\it HST\/} spectra of the intermediate and high ionization
species, especially the strong {\CIVdblt}, {\NVdblt}, and {\OVIdblt}
doublets, and of several other important low ionization species.

Presently, it is not clear if these low resolution data can be used to
place meaningful constraints on the chemical and ionization conditions
of the clouds in {\MgII} selected absorbers.  
In this paper, we investigated this issue in a pilot study, since an
affirmation would imply that a larger sample could be studied using
existing data from the {\it HST\/} Archive.
We would then be able to address the broader implications for galaxy
formation scenarios based upon the inferred metallicities, abundance
patterns, and inferred relative spatial distribution of the low and
high ionization absorbing
gas clouds.

\begin{figure*}[th]
\figurenum{1}
\plotfiddle{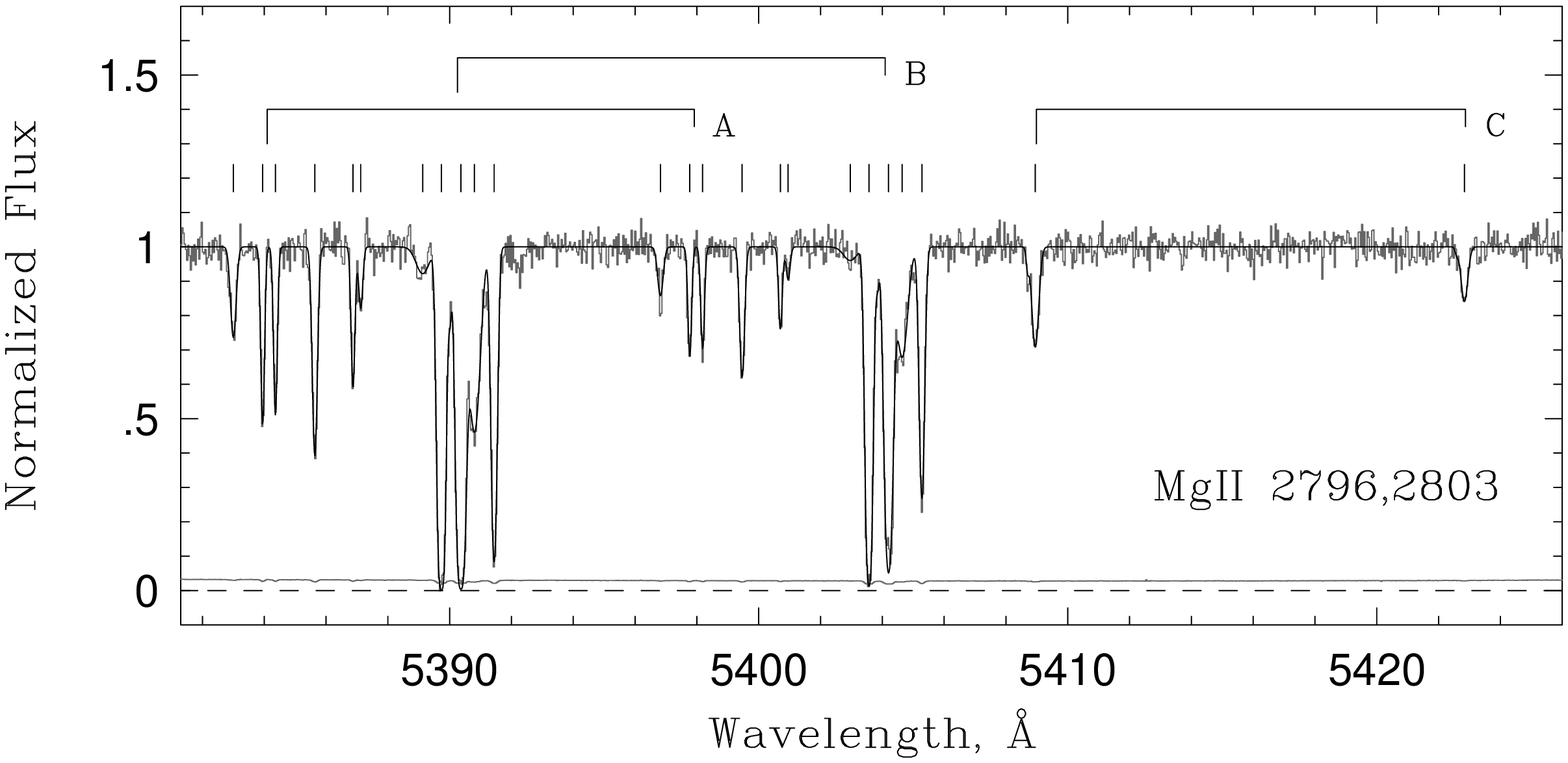}{2.8in}{0}{60}{60}{-236}{-74}
\protect\caption[Churchill.fig1.eps]
{\footnotesize
The HIRES/Keck {\MgIIdblt} doublet profiles at $R=45,000$.  Three
systems, A, B, and C, have been identified at  $z_{A} = 0.92540$,
$z_{B} = 0.92760$, and $z_{C} = 0.93428$.  The labeled bar bar ticks
give the systemic redshifted wavelengths for the {\MgII} doublets in
these systems. The solid line through the data is a synthetic spectrum
from a Voigt profile decomposition.  The vertical ticks above the
continuum give the locations of the Voigt profile
components. \label{fig:order28}}
\end{figure*}

Under the assumptions of photoionization and/or collisional ionization
equilibrium, we developed a technique in which it was assumed that
the number of clouds and their kinematics are obtained by Voigt profile
decomposition of the high resolution {\MgII} spectra.
We then used the lower resolution profiles from the FOS data to place
constraints on the range of chemical and ionization conditions in
these clouds.
We also explored the idea that the {\MgII} could arise in relatively
low ionization clouds embedded in a higher ionization and more
extended medium (see \cite{bergeron94}; \cite{iap}).
More specifically, we set out to answer three questions:
(1) Assuming the {\MgII} clouds measured with HIRES are photoionized,
can we construct model clouds that are consistent with the many
low and intermediate ionization species captured in the FOS data?  
(2) If so, are we required to infer an additional (presumedly low
density and diffuse) component to account for higher ionization
absorption from {\CIV}, {\NV}, and {\OVI}?
(3) If so, can this diffuse component be made consistent with
photoionized only, collisionally ionized only, or photo plus
collisionally ionized gas?

For this paper, we chose the three systems at $z _{\rm abs} =
0.9254$, $0.9276$, and $0.9343$ along the line of sight toward
PG~$1206+459$ ($z_{\rm em} = 1.16$) because they are exceptionally
rich in low, intermediate, and high ionization ultraviolet transitions
(\cite{bt96}; \cite{uvconf}; \cite{kp13}). 
The HIRES/Keck {\MgII} profiles are illustrated in
Figure~\ref{fig:order28}.
Two of the systems are kinematically ``complex'' and are separated by
$\sim 300$~{\kms}.
The third system is more isolated, being $\sim +1000$~{\kms} from the
other two.
This system is classified as a ``weak'' {\MgII} absorber [defined by
$W_{\rm r}(2796) < 0.3$~{\AA} (\cite{weakmgII})].
The highest ionization transitions, {\CIV}, {\NV} and {\OVI}, are
seen to have a total kinematic spread of $\sim 1000$~{\kms} coincident
with the three system seen in {\MgII} absorption (\cite{uvconf}).

In the QSO field, Kirhakos \etal (1994\nocite{kirhakos94}) found three
bright galaxies with angular separations from the quasar of 5.6, 8.6,
and 9.0{\arcsec} and $g$ magnitudes 21.1, 21.5, and 22.3,
respectively.
The  8.6{\arcsec} galaxy has detected {\OII} $\lambda 3727$ with flux
$9 \times 10^{-17}$ ergs {\cmsq} s$^{-1}$ at $z=0.93$ (\cite{thimm95}).
At this redshift, the QSO--galaxy impact parameters are $29$, $45$,
and $47~h^{-1}$~kpc ($q_0 = 0.05$).
There are $\sim 10$ galaxies with $21 \leq g \leq 22$ within
100{\arcsec} of the QSO (\cite{kirhakos94}).
This is an overdensity by a factor of $\sim 3$ compared to field
galaxies (\cite{tyson88}).
Thus, it is of interest to entertain the possibility of a group
environment for these absorbers.

In \S~\ref{sec:data} we describe the data and its analysis.
In \S~\ref{sec:models} we outline our modeling technique and
simplifying assumptions.
A synopsis of the model results are given in \S~\ref{sec:results}.
Details on how the data were used to constrain the models and how
various ionizing spectral energy distributions modify these models
are given in Appendices~\ref{app:constraints} and \ref{app:models}.
In \S~\ref{sec:discussion}, we compare and contrast the system
properties, and in \S~\ref{sec:onthe}, we discuss what might be
inferred about the relative spatial distribution of the low and high
ionization gas.
We summarize in \S~\ref{sec:conclusion}.

\section{Data and Analysis}
\label{sec:data}

\subsection{HIRES/Keck}

The optical data were obtained with the HIRES spectrometer
(\cite{vogt94}) on the Keck~I telescope on 23 January 1995 UT under
clear and stable conditions with a seeing of $\sim 0.6${\arcsec}.
The spectral resolution is $\sim 6.6$~{\kms} ($R=45,000$), with a
sampling of 3 pixels per resolution element.
The  signal--to--noise ratio is $\sim 50$ per resolution element.
The {\FeII} $\lambda 2344$, $2374$, $2383$, $2587$, and $2600$
transitions were captured at similar signal--to--noise ratio.

\begin{figure*}[th]
\figurenum{2}
\plotfiddle{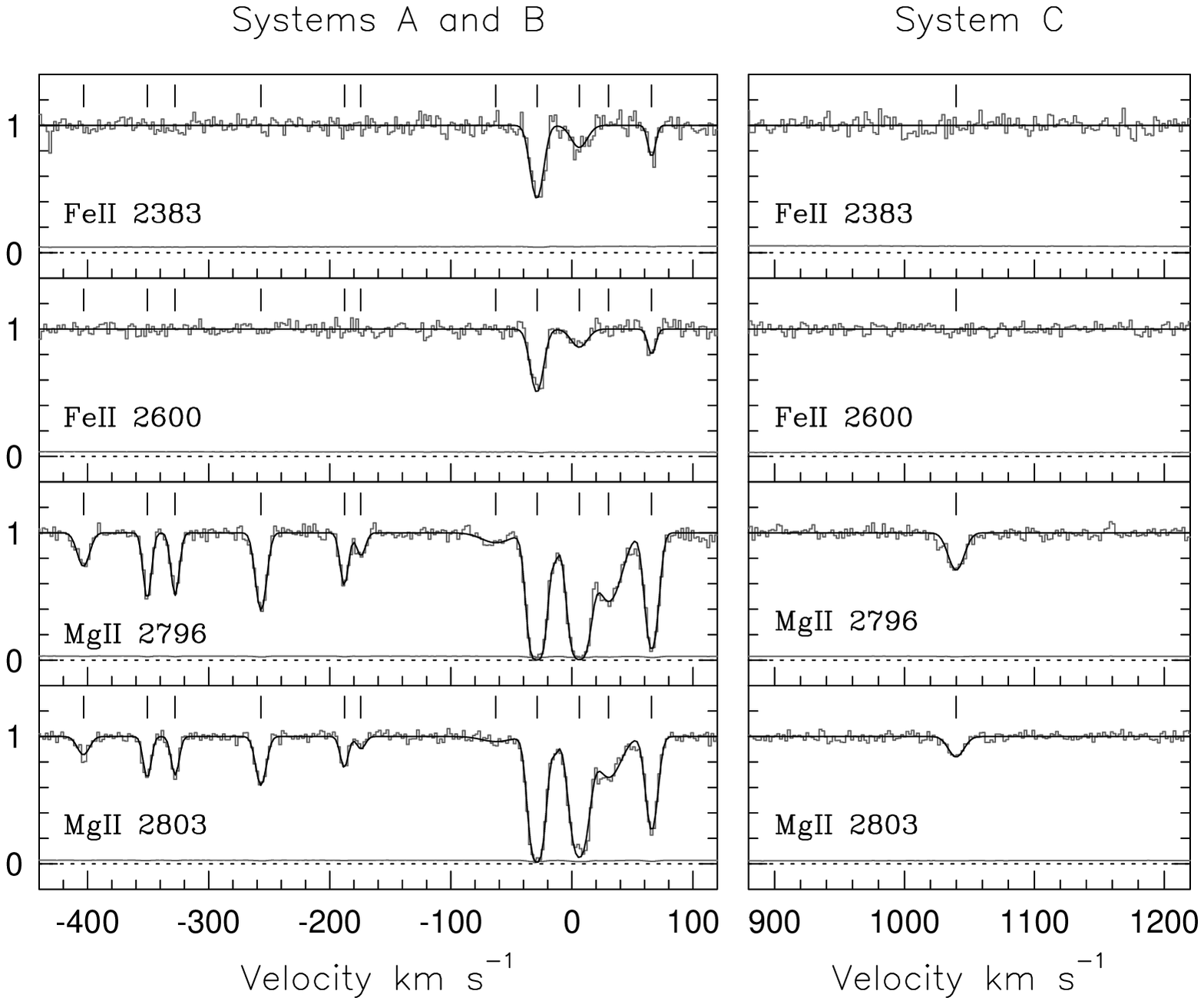}{3.9in}{0}{75}{75}{-236}{-274}
\protect\caption[Churchill.fig2.eps]
{\footnotesize
The kinematics of the {\MgII} systems are shown, with individual
clouds seen in {\FeII} $\lambda 2383$, {\FeII} $\lambda 2600$,
{\MgII} $\lambda 2796$, and {\MgII} $\lambda 2803$ absorption aligned
in rest--frame velocity.  Systems A and B are shown together in the
left panel and system C is shown in the right panel.  The zero--point
is defined at $z=0.92760$. The solid lines through the data are
synthetic spectra from a Voigt profile decomposition.  The ticks above
the continuum give the locations of the Voigt profile components.
\label{fig:vpfits}}
\end{figure*}

The HIRES spectrum was reduced with the IRAF\footnote{IRAF is
distributed by the National Optical Astronomy Observatories, which are
operated by AURA, Inc., under contract to the NSF.} {\it Apextract\/}
package for echelle data.  The detailed steps for the reduction are
outlined in Churchill (1995\nocite{lotr}).
The spectrum was extracted using the optimized routines of Horne
(1986\nocite{horne86}) and Marsh (1989\nocite{marsh89}). 
The wavelengths were calibrated to vacuum using the IRAF task {\it
ecidentify\/}, which models the full 2D echelle format.
The absolute wavelength scale was then corrected to heliocentric
velocity.
The continuum normalization was performed using the IRAF {\it sfit\/}
task.
Objective and unbiased identification of absorption features (without
regard to their association with the studied systems) was performed as
described in Churchill \etal (1999b\nocite{cvc99}), using the
methodology of Schneider \etal (1993\nocite{schneider93}).

Three systems, which we hereafter call A, B, and C, are observed at
redshifts $z_{A} = 0.92540$, $z_{B} = 0.92760$, and $z_{C} = 0.93428$.
In Figure~\ref{fig:order28}, we present the {\MgII} doublet profiles.
The doublets are marked by the labeled bar above the normalized
continuum.
The solid curve through the data is a model spectrum generated using 
Voigt profile (VP) decomposition.
The free parameters are the number of VP components (``clouds'') and,
for each, its redshift, column density, and Doppler $b$ parameter.
We have used the program MINFIT (\cite{cwcthesis}), which performs a
$\chi^{2}$ minimization while minimizing the number of clouds using 
a specified confidence level and the standard F--test.
The adopted VP decomposition had six clouds in system A, five clouds
in system B, and a single cloud in system C.

The HIRES data and the VP decompositions are shown in
Figure~\ref{fig:vpfits} with the profiles aligned in rest--frame
velocity. 
In Table~\ref{tab:vpfits}, we present the cloud properties, including
individual redshifts, velocities with respect to $z = 0.92760$, column
densities, $b$ parameters, and the ratios $\log \{
N({\FeII})/N({\MgII}) \} $.
Only system B was found to have measurable {\FeII}.
The $1\sigma$ upper limits on {\FeII} were obtained for each cloud
from the equivalent width limits of the $\lambda 2600$ transition.  
For {\FeII} in system A, we measured a mean $1\sigma$ column density
upper limit of $N({\FeII}) \leq 10^{11.1}$~{\cmsq} for the six clouds 
using the technique of stacking (\cite{norris83}).

\phantom{dummy space.}

\subsection{FOS/{\it HST\/}}

The ultraviolet data were obtained with the Faint Object Spectrograph
(FOS) on the {\it Hubble Space Telescope\/} as part of the QSO
Absorption Line Key Project.
The data acquisition, their reduction, and the objective absorption
line lists are presented in Jannuzi \etal (1998\nocite{kp13}).
Their fully reduced G190H and G270H spectra have kindly been made
available for this study.
The spectra have a resolution of $R=1300$ ($\sim 230$~{\kms}) and cover
the approximate wavelength intervals $1600$ to $2313$~{\AA} and $2225$
to $3280$~{\AA} for the G190H and G270H settings, respectively.

For the most part, we have adopted the Key Project continuum fits and
line identifications.
Some refinement was needed near the {\SiIVdblt} doublet ($2680 \leq
\lambda _{\rm obs}  \leq 2715$~{\AA} on G270H), which lies on the red
wing of the broad {\Lya} emission line, and near the {\SiIIdblt}
doublet ($2290 \leq \lambda _{\rm obs} \leq 2310$~{\AA} on G190H),
which lies at the spectrum edge.
We have re--fit the continuum across these regions.

A Lyman limit break is present at $\sim 1760$~{\AA} (however, see
Stengler--Larrea 1995\nocite{kp5}; Jannuzi\nocite{kp13} et~al.\ 1998).
We extrapolated the continuum fit of Jannuzi \etal below the break
starting at the break shoulder ($\lambda = 1895$~{\AA}).
This technique preserved the measured optical depth, or the break
ratio, $F_{+}/F_{-}$ (\cite{schneider93}), while yielding a reasonable
approximation to the shape of the recovery (see
Figure~3$a$).
Due to the high density of lines, the Jannuzi \etal continuum may have
been systematically low by 5--10\% in the region 1790 to 1820~{\AA}
and our extrapolation may have propagated this systematic offset.
Nonetheless, this error has a negligible effect on the measured break
ratio of $2.5\pm0.4$, which we obtained from the unnormalized as well
as the normalized spectrum. 
This ratio implies $\tau_{\rm LL} \simeq 0.9$ and a total neutral
hydrogen column density of $N({\HI}) \simeq 10^{17.2}$~{\cmsq}.

Out of the $\sim 70$ QSOs analyzed by the QSO Absorption Line Key Project,
PG~$1206+459$ is one of eight for which the line identifications
(IDs) were subject to greater uncertainty (the {\Lya}
forest is ubiquitous blueward of 2623~{\AA} and at least four metal
systems are present).
Using the HIRES data for cross checks with the Jannuzi\nocite{kp13}
\etal  line IDs in the FOS spectrum, we have made a table of
transitions, including those detected and those used for constraining
the models. 
In Table~\ref{tab:fosIDs} (end of paper), we have listed the observed
wavelengths, line IDs, ionization potentials, and notes on any
blending. 
For example, we identified the Lyman series down to {\Lye}, beyond
which the three systems blend together. 
The FOS spectrum provides a large number of chemical and ionization
species, including
{\Lya}, {\Lyb}, {\Lyg}, {\Lyd}, {\Lye}, {\CII} $\lambda 1036$ and
$\lambda 1334$, {\CIII} $\lambda 977$, {\CIVdblt}, {\NII} $\lambda
915$, {\NIII} $\lambda 989$, {\NVdblt}, {\OVIdblt}, {\SiIIdblt} and
$\lambda 1260$, {\SiIII} $\lambda 1206$, {\SiIVdblt}, and {\SVIdblt}.
These species represent a wide range of ionization potentials from 
13.6~eV for {\HI} to 138.1~eV for {\OVI}.


\section{The Models}
\label{sec:models}

Assuming the {\MgII} clouds are photoionized, our goal was to
determine if we can obtain useful constraints on their range of chemical
and ionization conditions using the FOS/{\it HST\/} spectra.
Within this context, we also explored the possibility of a high
ionization ({\CIV}, {\NV}, and {\OVI}), presumedly diffuse, component.
In principle, this high ionization phase could be photoionized,
collisionally ionized, or photo plus collisionally ionized gas (a
spatially segregated two--phase high ionization component).

For both the {\MgII} clouds and the high ionization phase, the
extragalactic ultraviolet ionizing background spectrum of Haardt
\& Madau (1996\nocite{handm96}) for $z=1$ was assumed.
Using the {\OII}~$\lambda 3727$ detection and constraints measured by
Thimm (1995\nocite{thimm95}), we explored the range of allowed
contributions (modifications to the Haardt \& Madau background) from
various galactic spectral energy distributions.
We find that galactic contributions, within the allowed ranges
explored, do not modify {\it general\/} conclusions based upon the
assumption of a pure Haardt \& Madau background (see
Appendix~\ref{app:models}).

\subsection{The {\MgII} Clouds}

We used the photoionization code CLOUDY (version 90.4;
\cite{ferland}).
The free parameters are the neutral hydrogen column density,
$N({\HI})$, the metallicity\footnote{In this work we use the notation
${\rm [X/Y]} = \log ({\rm X/Y}) - \log ({\rm X/Y})_{\odot}$ and for
metallicity use $Z \equiv \log Z/Z_{\odot}$.}, $Z$, the abundance
pattern, and the ionization ``parameter'', $U$, which is defined as
the ratio of the number of hydrogen ionizing photons to the hydrogen
number density, $n_H$ (including ionized, neutral, and molecular
forms). 
For the Haardt \& Madau spectrum and  normalization, a simple relation
between ionization parameter and hydrogen number density, $\log U =
-5.2 - \log n_{H}$, holds.

For a given abundance pattern, the ratio $N({\FeII})/N({\MgII})$
uniquely determines the ionization parameter, $U$.
Once $U$ is determined, the measured $N({\MgII})$ fixes $\log
N({\HI}) + Z \simeq C_{1}$, for a cloud in photoionization
equilibrium.
For a given $Z$, the constant $C_{1}$ is constrained by the Lyman
series transitions in the FOS data.
Throughout, we assume $Z \leq 0$.
To characterize the abundance pattern, we used the ratio of
$\alpha$--group species to Fe--group species, [$\alpha$/Fe].
Abundance ratios measured in Galactic stars show a clear range of 
$0 \leq [\alpha/{\rm Fe}] \leq +0.5$ (\cite{jtl96}), which we adopt
as a reasonable range for the studied systems.
Since all clouds have measured $N({\MgII})$, an $\alpha$--group
element, it follows that $[\alpha/{\rm Fe}] + Z \simeq C_{2}$.
Thus, for clouds with measured $N({\FeII})/N({\MgII})$, the only
arbitrarily chosen free parameter is the abundance pattern.  
One selects a $Z$ and [$\alpha$/Fe] (fixes $C_{2}$), determines
$N({\HI})$ by constraining the photoionization models with the Lyman
series transitions, and thus determines $C_{1}$ (an example of
this process is given in Appendix~\ref{app:constraints}). 

By exploring a large range of $Z$ and [$\alpha$/Fe], we verified the
above relationships.  
When only an upper limit is available on $N({\FeII})/N({\MgII})$ for a
given cloud, one has two arbitrarily chosen parameters, [$\alpha$/Fe]
and $N({\FeII})/N({\MgII})$, which together uniquely determine the
ionization parameter.
The constraints on the {\MgII} cloud ionization parameters, abundance
pattern, and metallicities are fairly tight and robust; even when 
$N({\FeII})/N({\MgII})$ was an upper limit, the {\SiII}, {\SiIII}, and
{\SiIV} ratios were key for constraining the ionization parameter,
independent of the abundance pattern (see
Appendix~\ref{app:constraints}).
In a given system, for clouds in which $N({\FeII})/N({\MgII})$ was
only an upper limit, we assumed that they were identical
vis--\'{a}--vis their {\MgII} column densities (had identical
metallicities and abundance patterns).
This yielded model clouds with identical $Z$, and [$\alpha$/Fe], but
unique $N({\HI})$, due to their unique $N({\MgII})$.
The allowed range of cloud to cloud variations within a system, if
desired, can be obtained from the two relations giving $C_{1}$ and
$C_{2}$ (as long as the total $N({\HI})$ is held constant).

To narrow parameter space, we began with a grid of photoionization
models with CLOUDY, where the grid was defined for  
(1) $\log \{ N({\FeII}) / N({\MgII}) \} $ from $-1$ to $-4$ in
intervals of 0.5 dex (this provides the ionization parameter, $U$),
(2) $N({\HI})$ from $10^{14}$ to $10^{18}$~{\cmsq} in 1 dex intervals,
(3) $Z$, from $-2.0$ to $+0.4$ in intervals of 0.2 dex, and
(4) solar and $[\alpha/{\rm Fe}] = +0.5$ abundance pattern.
Once the parameter space was narrowed, we ran CLOUDY in its optimized
mode tuned to the {\MgII} column densities, to obtain the adopted
models.

\begin{figure*}[ht]
\plotfiddle{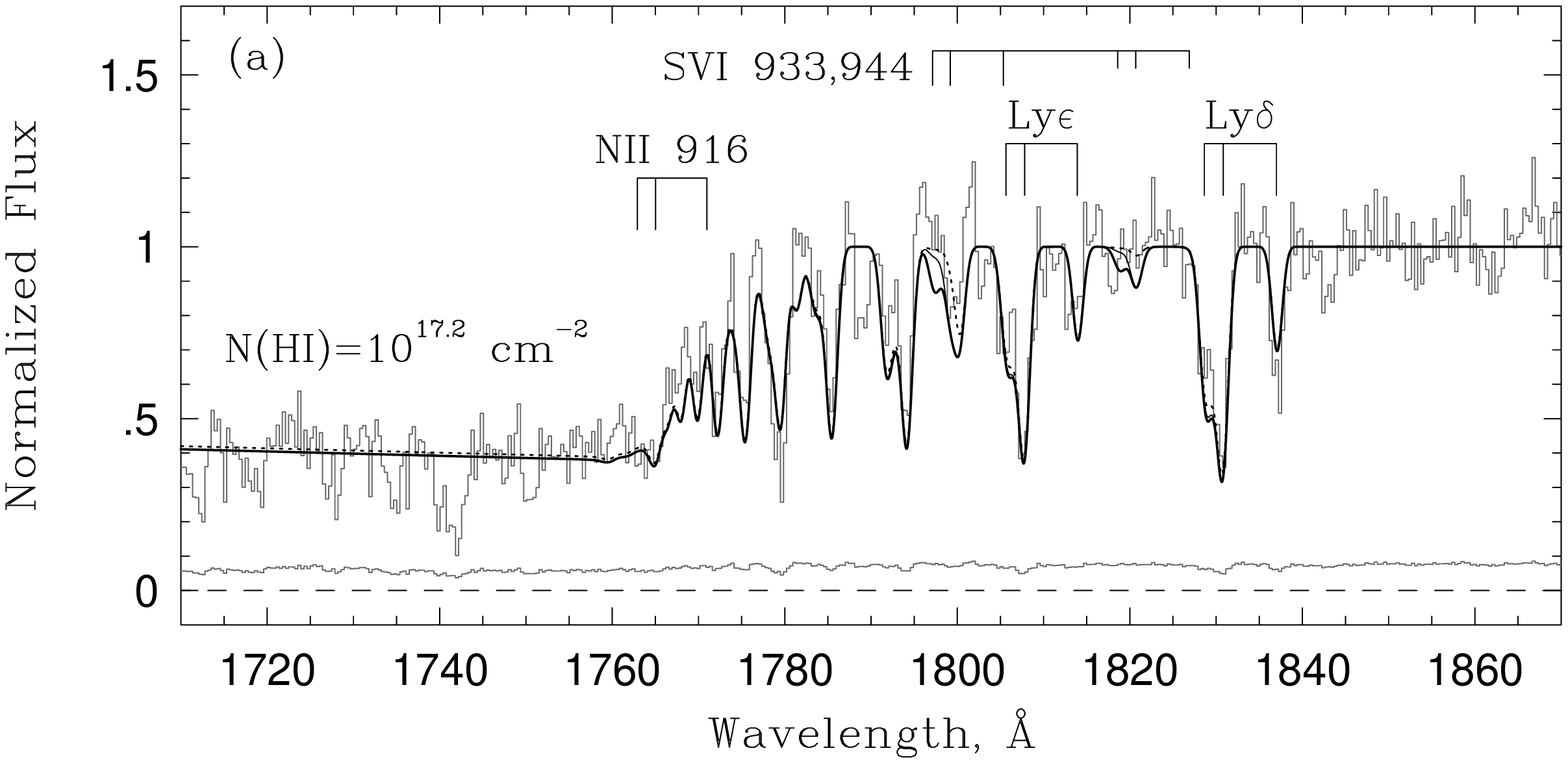}{5.5in}{0}{60}{60}{-236}{121}
\plotfiddle{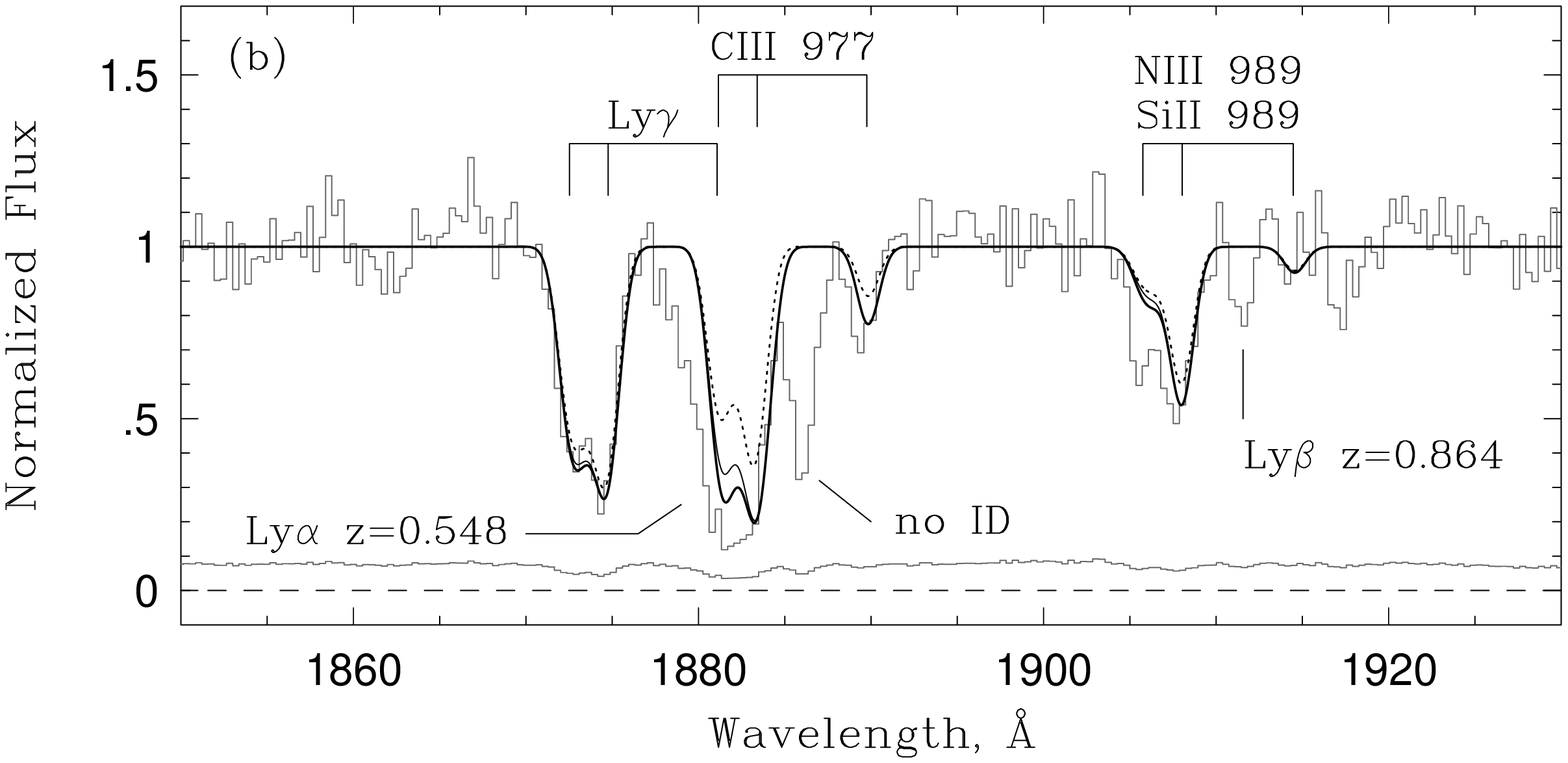}{0.in}{0}{60}{60}{-236}{-74}
\figurenum{3$a$--$b$}
\protect\caption
{\footnotesize
The normalized FOS/{\it HST\/} spectrum (histogram) and the tuned model
predictions (not fits).  The ticks mark locations of the constraint
transitions identified for systems A, B, and C.  Three models are
shown. The thick solid spectrum includes the photoionized {\MgII}
clouds, the photoionized diffuse component for systems B and C, and the
photo plus collisionally ionized diffuse component for system
A.  The narrow solid spectrum includes the {\MgII} clouds and the
photoionized diffuse components for each system.  The dotted--line model
includes the {\MgII} clouds only and is shown to illustrate the
required high ionization phases.
---($a$) The Lyman break and Lyman series.  See text regarding the
continuum fit, which is probably low in the region of 1760~{\AA} to
1820~{\AA}.  Note the {\NII} $\lambda 916$ in system B and the
{\SVIdblt} in systems A and B.
---($b$) The {\Lyg}, {\CIII} $\lambda 977$ and {\NIII} $\lambda 989$
predictions.  Note the sensitivity of {\CIII} to the model components.
}
\end{figure*}

\begin{figure*}[ht]
\figurenum{3$c$--$d$}
\plotfiddle{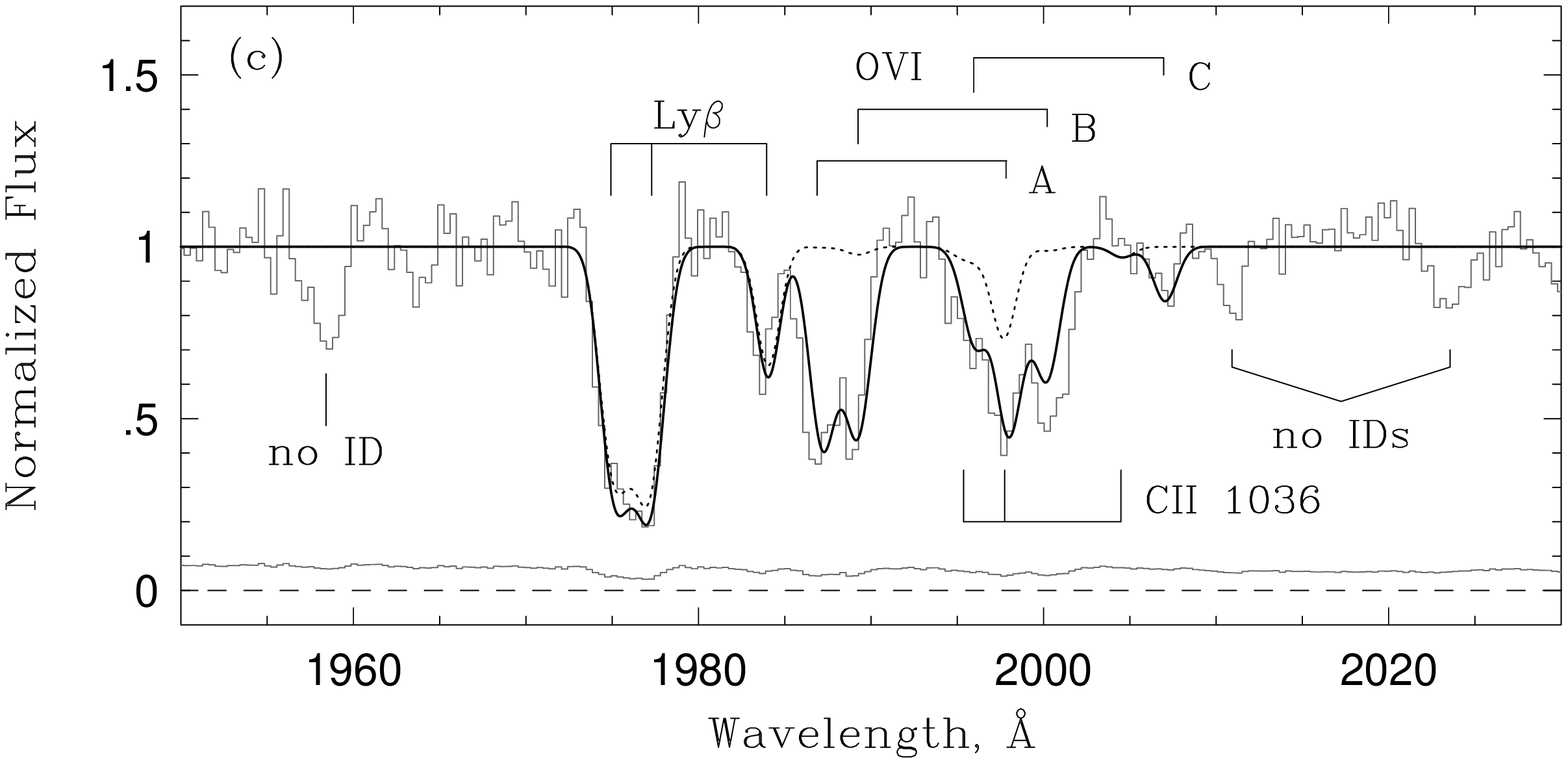}{5.5in}{0}{60}{60}{-236}{121}
\plotfiddle{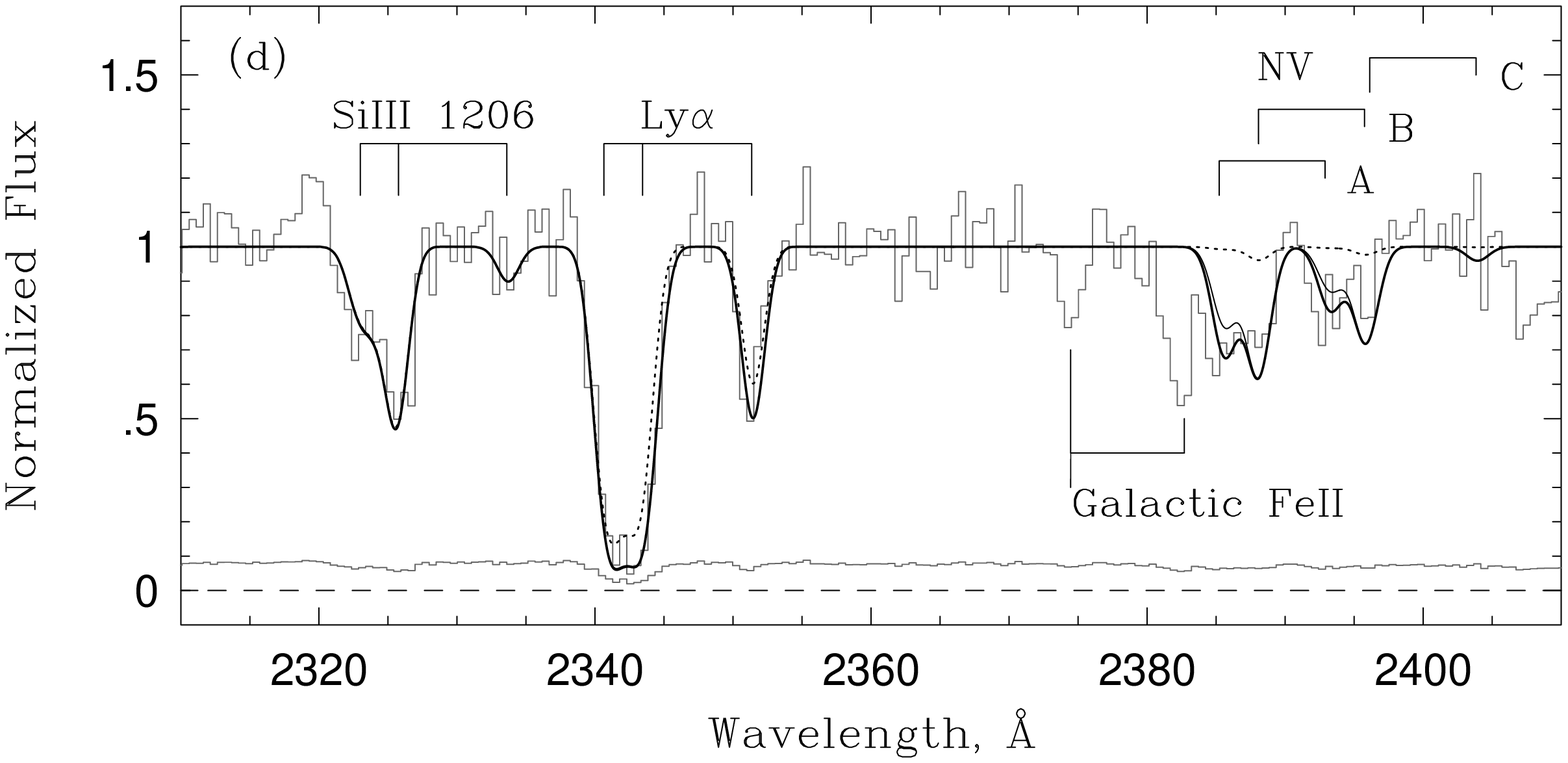}{0.0in}{0}{60}{60}{-236}{-74}
\protect\caption
{\footnotesize
Same as Figure 3$a$ and 3$b$.
---($c$) The {\Lyb}, {\OVIdblt}, and {\CII} $\lambda 1036$ predictions.
The wavelength calibration may be shifted blueward in the region
$\sim 1980$ to 2005~{\AA}.  Virtually no {\OVI} resides in the {\MgII}
clouds, but must arise in a higher ionization phase.
---($d$) The {\SiIII} $\lambda 1206$, {\Lya}, and {\NVdblt} predictions.
Note the sensitivity of {\Lya} and {\NV} to the model components.  The
{\SiIII} arises primarily in the {\MgII} clouds.
}
\end{figure*}

\begin{figure*}[ht]
\figurenum{3$e$--$f$}
\plotfiddle{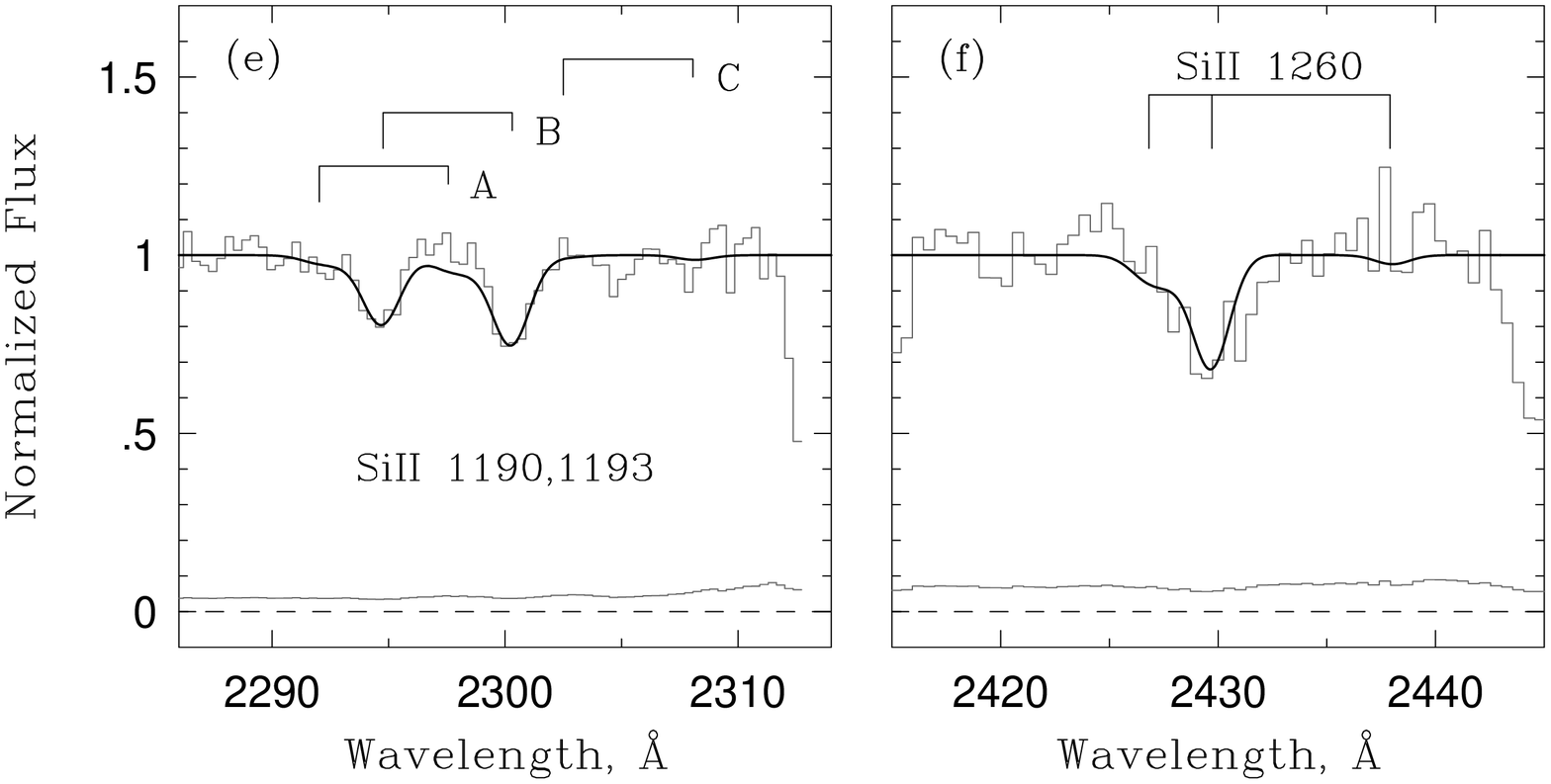}{5.5in}{0}{60}{60}{-236}{121}
\plotfiddle{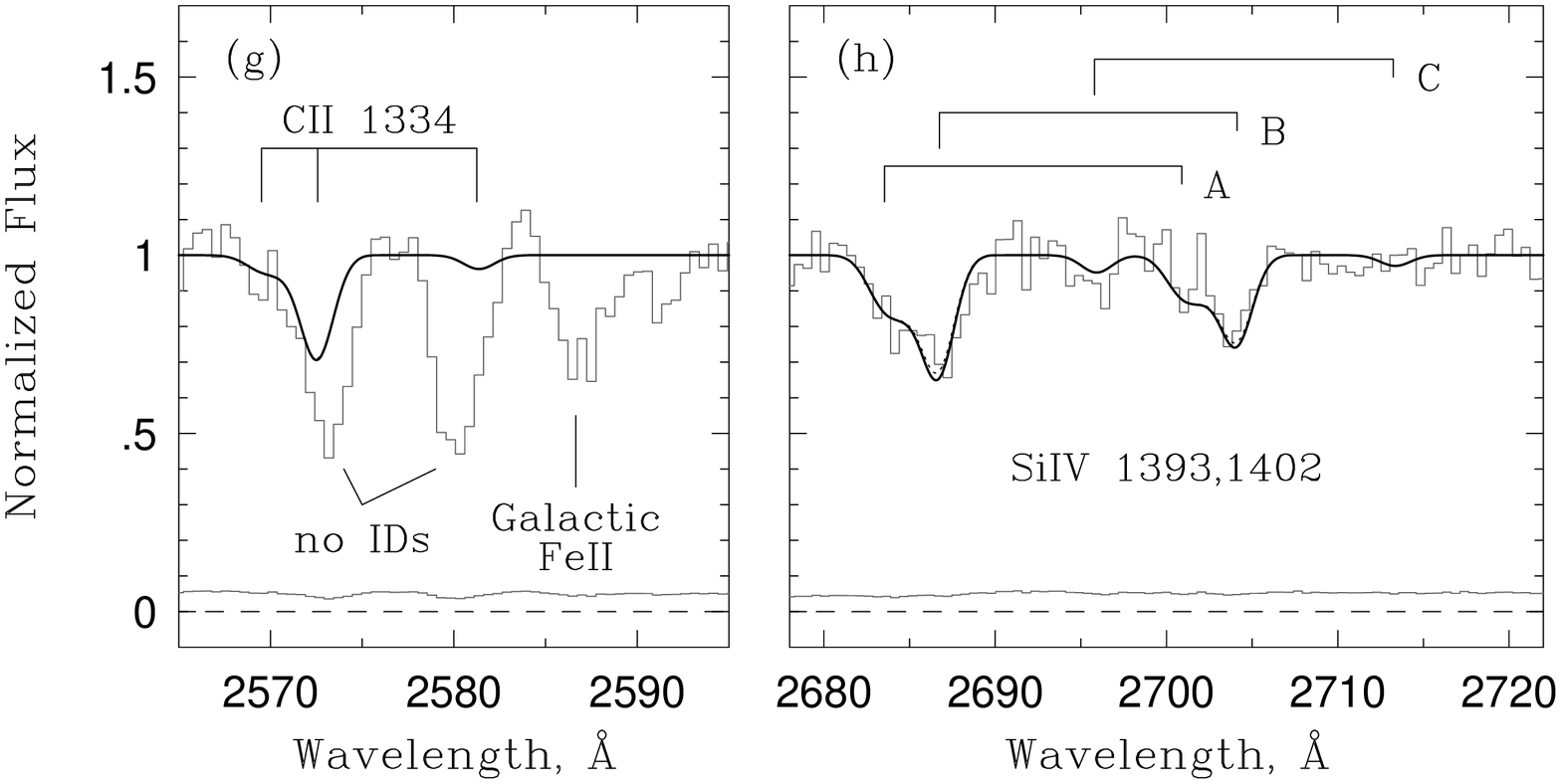}{0.0in}{0}{60}{60}{-236}{-74}
\protect\caption
{\footnotesize
Same as Figure 3$a$ and 3$b$.
---($e$) The {\SiIIdblt} doublet.  The {\SiII} also arises primarily
in the {\MgII} clouds.
---($f$) The {\SiII} $\lambda 1260$ prediction.  A blend ({\Lya}?) must
be present in the red wing.
---($g$) The {\CII} $\lambda 1334$ prediction.  As with {\SiII}
and {\SiIII}, the {\CII} arises primarily in the {\MgII} clouds.  Note
the strong blend ({\Lya}?) in the red wing.
---($h$) The {\SiIVdblt} doublet predictions.  Since the {\SiIV} also 
arises primarily in the {\MgII} clouds, the {\SiII}, {\SiIII} and
{\SiIV} ratios placed tight constraints on the cloud ionization
conditions.
}
\end{figure*}

\begin{figure*}[ht]
\figurenum{3$i$}
\plotfiddle{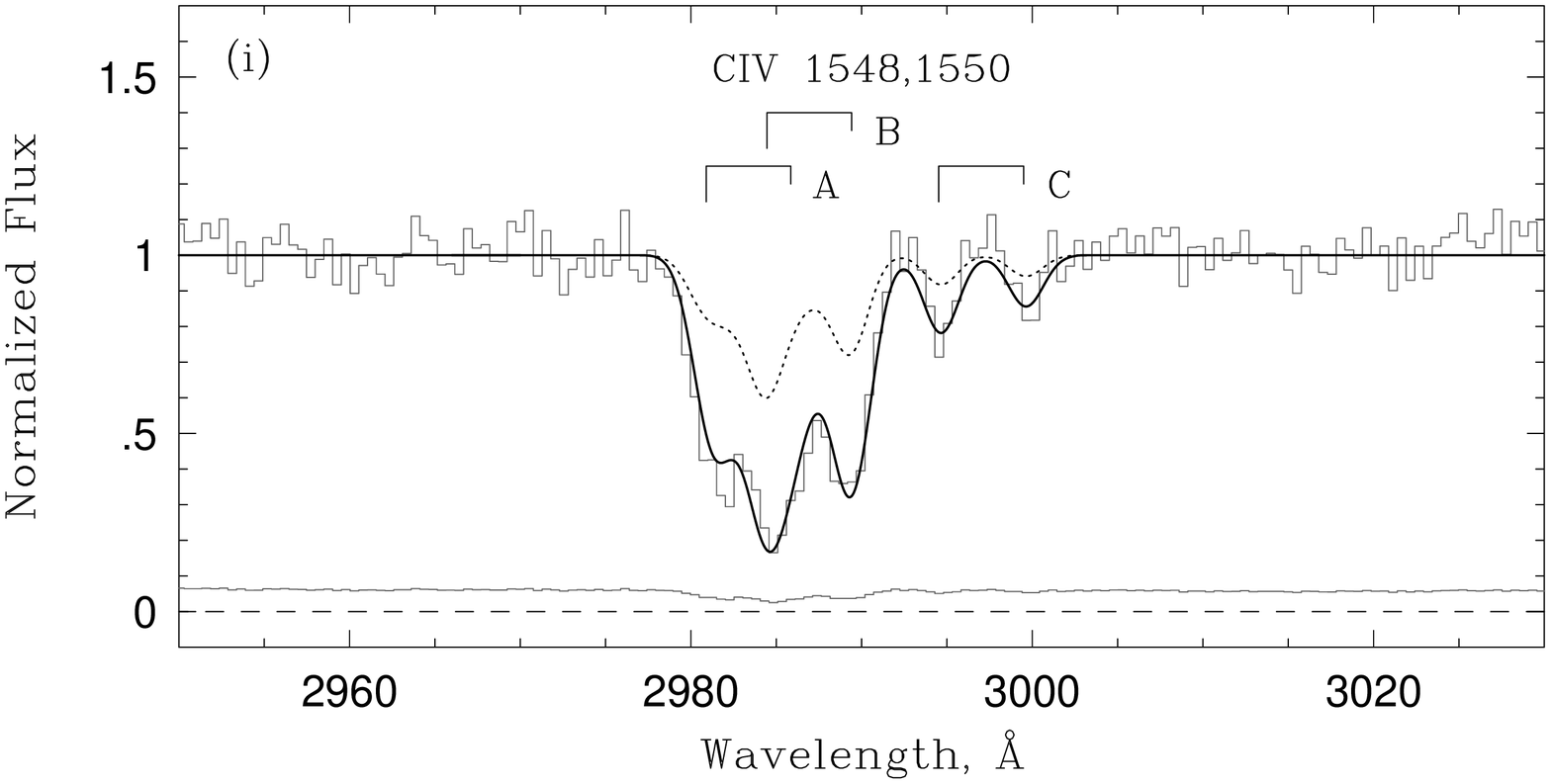}{2.82in}{0}{60}{60}{-236}{-70}
\protect\caption
{\footnotesize
Same as Figure 3$a$ and 3$b$.
The (self--blending) {\CIVdblt} doublet predictions.  For
systems A and B, note that a fair fraction of the {\CIV} arises in the
{\MgII} clouds, but the majority must arise in a higher ionization
phase. \label{fig:3i} \label{fig:figure3}}
\end{figure*}

Model clouds are drawn from the grid and a synthetic FOS spectrum is
generated from the model column densities for the transitions listed
in Table~\ref{tab:fosIDs} (end of paper).
The simulated FOS spectrum is generated by modeling the absorption
from each transition by Voigt profiles convolved with the FOS
instrumental spread function.
The Doppler parameters of the modeled transitions are determined from
the observed {\MgII} $b$ parameters and the kinetic temperature, $T$,
output by the CLOUDY models, typically 5000 to 30,000~K.  
The CLOUDY temperature is used to estimate the thermal component of
$b({\MgII})$, from which the turbulent $b$ parameter is computed
from the relation, $b_{tot}^{2} = b_{thermal}^{2} + b_{turb}^{2}$.
The total $b$ parameter of any transition can then be estimated from
its CLOUDY thermal $b$ and the turbulent $b$.
The synthetic spectrum is then superimposed on the observed FOS
spectrum and the $\chi^{2}$ is calculated pixel by pixel in regions of
interest as an indicator of the goodness of the model.

\subsection{High Ionization Gas}

As will be shown, the model {\MgII} clouds could not account for even
a small fraction of the absorption strengths of the higher ionization
species (i.e.\ {\CIV}, {\SVI}, {\NV}, and {\OVI}).  
Thus, we postulated a high ionization diffuse component not seen
in {\MgII} absorption.
A maximal diffuse scenario provides the low ionization limit of the
{\MgII} clouds such that their contribution to the moderate and high
ionization species is minimized.
A minimal diffuse scenario provides the high ionization limit of the 
{\MgII} clouds such that their contribution to the moderate and high
ionization species is maximized.

For a postulated high ionization component, we used {\CIV} and {\OVI}
to constrain model cloud properties for both a solar abundance
pattern, $\hbox{[C/O]} = 0$, and an oxygen to carbon enhancement of
$\hbox{[C/O]} = -0.5$.
For constraints on the models from {\SVI}, we use $[{\rm S}/{\rm H}] =
0$, but note that sulfur is enhanced by $\sim 0.5$ dex for $[{\rm
Fe}/{\rm H}] < 0$.
We avoid using {\NV} as a primary constraint because the chemical
enrichment processes for nitrogen can lead to wide variations in its
abundance (\cite{wheeler89}).
We adopt $\hbox{[N/O]} = 0$ for this work, but occasionally discuss
possible variations in this ratio.

To examine the range of properties of a diffuse component, we did the
following systematic explorations.
For each system, we first assumed that any required high ionization
gas is in a {\it single}, broad component.
The adopted models of the high ionization components are obtained
using the same techniques illustrated in Appendix~\ref{app:constraints},
but the primary constraints are the residual strengths of the {\CIV}
and {\OVI} absorption unaccounted by the {\MgII} clouds.
Also, consistency with any unaccounted {\CIII}, {\SiIII}, {\SiIV},
{\SVI} and {\NV} is required.
We then explored the possibility that the {\CIV} arises in a few
``{\CIV}--only'' clouds while the {\OVI} arises in a diffuse low
density high ionization phase that is separate from the {\CIV} clouds,
which are separate from the {\MgII} clouds.
The {\NV} and {\SVI} absorption profiles were critical for
constraining these ``{\CIV} cloud'' models.

We systematically explored if contributions from both photoionization
and collisional ionization are consistent with the data. 
For collisional ionization, we have drawn from the equilibrium models
of Sutherland \& Dopita (1993\nocite{sutherland93}), with solar
abundances.
In these models, the relative column densities of various species,
including that of neutral hydrogen, are a unique function of
temperature for a given metallicity.  
Thus, under the model assumptions, the remaining free parameter is the
temperature.
To do so, we located the lowest ionization level of a photoionized
diffuse component consistent with the intermediate and moderately high
ionization species, after taking into account the {\MgII} clouds.
A collisional ionization model  was then tuned to any unaccounted
absorption in {\NV} and {\OVI}.


\section{Model Results}
\label{sec:results}

Our intent is to present a general picture of both the {\MgII} cloud
and high ionization diffuse component properties within the context of
our modeling.
The adopted model properties are presented in Tables~\ref{tab:cloudpars}, 
\ref{tab:PICs}, and \ref{tab:DICs}.  
In Table~\ref{tab:cloudpars}, we list the physical parameters of the
twelve {\MgII} clouds for the scenarios of a maximal or minimal diffuse
component. 
Typical model parameters for each system are given in
Table~\ref{tab:PICs}.
The tabulated values serve as a guide; in the following discussion we
quote allowed ranges in the cloud properties based upon 
\S~\ref{sec:models}.
The diffuse component properties are listed in Table~\ref{tab:DICs}.
The results described below are for the Haardt \& Madau
(1996\nocite{handm96}) extragalactic background.
In Appendix~\ref{app:models}, we describe our explorations with several
galactic/starburst radiation fields (representing both somewhat
extreme and ``normal'' cases) and argue that our general conclusions
are not altered.
In Figure~3, we present the synthetic spectrum of our
models superimposed upon the FOS spectrum.
The model transitions are labeled with three--point ticks, which
give the locations of systems A, B and C, from blue to red,
respectively.
Three synthetic spectra are shown.
The dotted--line spectrum is of the twelve photoionized {\MgII}
clouds.
The thin solid--line spectrum includes both the {\MgII} clouds and 
the single--phase photoionized diffuse components.
The thick solid--line spectrum includes a two--phase photo plus
collisionally ionized component in system A.

\subsection{System A $(z=0.9254)$}

Since $N({\FeII})/N({\MgII})$ was not measured in any of the system A
clouds, the clouds were assumed identical vis--\'{a}--vis their
$N({\MgII})$.
The {\SiII} to {\SiIV} ratio tightly constrains the ionization level
of these six clouds.
There is only a very small range of allowed ionization conditions,
with $\log U \simeq -2.3$.
To avoid super solar metallicity, the mean $[\alpha / {\rm Fe}]$
ranges from $+0.3$ to $+0.5$, which corresponds to $0.0 \leq Z \leq
-0.2$, respectively, for the best match to the Lyman series.
We obtained $10^{14.9} \leq N({\HI}) \leq 10^{15.6}$~{\cmsq} in each
of the six clouds; system A makes a negligible contribution to the
Lyman limit break.

A highly ionized phase, not seen in {\MgII}, is required
to account for the observed {\CIV}, {\NV}, and {\OVI} absorption.
For a single component, a Doppler width of $50 \leq b \leq 100$~{\kms}
is consistent with the data.
We ran a series of simulations, focused on the blended {\CIV} profile,
to explore the best combination of Doppler widths for {\CIV} and {\OVI}
in systems A and B, and to determine the allowed ranges for fits with
with $\chi ^{2}_{\nu} \leq 2$.
Based upon these simulations, we adopted $b=70$~{\kms}.
This component is consistent with either a single--phase photoionized
diffuse medium with $\log U \simeq -1.2$, or a two--phase photo plus
collisionally ionized diffuse medium with $\log U \simeq -1.26$ and
$T\simeq 3\times 10^{5}$~K, respectively.
All phases must be metal rich, $Z\sim 0$, in order to not overproduce
the Lyman series.
It is not possible to establish which of the two provides a better
description of the high ionization transitions.
The latter yields a better match to the {\NV} absorption.
However, $[{\rm N}/{\rm O}] \simeq +0.15$ in the photoionized only
phase could alternatively explain the {\NV} absorption.
If the collisional component is present, further tests showed that
{\CIV} arising in a few clouds with smaller $b$ parameters was not
ruled out.
The bottom line is that a highly ionized phase is required, whether
the {\CIV} arises in a single or multiple clouds and the {\OVI} and
{\NV} in a collisionally ionized phase, or whether the {\CIV}, {\NV},
and {\OVI} all arise in a single highly photoionized phase.

For the photoionization scenario, the diffuse cloud size increased as
$b$ was decreased from $70$~{\kms}, primarily due to the curve of
growth behavior of {\CIV}; its column density increases rapidly with
decreasing $b$ and is effectively constant for $b>70$~{\kms} in the
relevant range of equivalent width.
Based upon exploration of the allowed range of $b$ parameters, an
upper limit on the size of the diffuse component is $\sim 30$~kpc for
$b = 50$~{\kms}.
However, the best models, incorporating both a $\chi^{2}_{\nu}$ fit to
the {\CIV} profile and matching to the {\NV}, {\SVI}, and {\OVI}
profiles, had $b \geq 65$~{\kms}, which yielded sizes of
$\sim 10$--$20$~kpc.

\phantom{dummy space.}

\phantom{dummy space.}

\subsection{System B $(z=0.9276)$}

A range of ionization conditions ($-3.2 \leq \log U \leq -2.7$) for
the clouds are found for this system, primarily based upon the
observed {\FeII} to {\MgII} ratios in clouds 8, 9, and 11, and upon
{\SiII}, {\SiIV} and the other low ionization species for clouds 7 and
10.
However, the low ionization limits (maximal diffuse scenario) are
ruled out because the {\SiIV}, {\NIII}, and a large fraction of the
{\CIII} must arise in the {\MgII} clouds.
The Lyman limit break is primarily produced by clouds 8 and 11.
For these clouds, the metallicity must be $Z \geq -0.2$, even for an
$\alpha$--group enhanced abundance pattern, or else the Lyman limit
break would be too large.
In clouds 7 and 10,  the allowed abundance pattern range is 
$ 0.0 \leq [\alpha/{\rm Fe}] \leq +0.5$, corresponding to $ -0.1 \geq
Z \geq -0.6$ (the lower limit is constrained by the Lyman series).
The {\SiIV} arises primarily in cloud 10, the most highly ionized,
lowest density, and extended ($\sim 5$~kpc) {\MgII} cloud of the five.
Even for the minimal diffuse scenario (high ionization limit), the
predicted {\CIV} absorption is still well below the observed strength.

Thus, an additional high ionization component, not seen in
{\MgII} absorption, is required to account for the {\CIV},
{\NV}, and {\OVI} absorption.
This component is consistent with a single--phase photoionized diffuse
medium ($b \simeq 70$~{\kms}, based upon the simulations of the {\CIV}
profile) with a solar abundance pattern and near solar metallicity.
The bulk of the {\CIV}, {\NV}, and {\OVI} absorption
arises in this component, whereas the {\NIII} and {\SiIV} absorption
arise primarily in cloud 10.
The {\CIII} arises in both the {\MgII} clouds and the diffuse
component.
In Figure~3$a$, we point out the {\NII} $\lambda 916$
absorption, which accounts for the reduced flux in the Lyman break.
We also point out the weak {\SVIdblt} absorption, which arises in the
diffuse component (the $\lambda 933$ transition is blended with
{\Lysix} from system C).
It is not possible for a collisionally ionized phase to
substantially contribute.

As with system A, we obtained a maximum line of sight size, $S$, of
$30$~kpc for the highly ionized diffuse component, but found a
preferred size range of $10$--$20$~kpc.
For $b \leq 50$~{\kms}, the larger cloud size elevated absorption in
all high ionization species such that, in particular, {\SVI} was
significantly overproduced.
A reduced $[{\rm S}/{\rm H}]$ would yield reduced sulfur absorption,
but is unlikely because sulfur is usually enhanced relative to solar
and is known to not suffer dust depletion (\cite{jtl96}).

\subsection{System C $(z=0.9343)$}

Though system C is a single weak {\MgII} cloud, it is best described
by two photoionized phases.
This inference is based upon a self--consistent match to the Lyman
series, which is obtained when both a narrow and a broad component are
included.
The narrower low ionization {\MgII} cloud accounts for the {\HI} in
the  {\Lyg}, {\Lyd}, and {\Lye} absorption, whereas the broader high
ionization diffuse component contributes significantly to the {\Lya}
and {\Lyb} profiles.

The lower ionization phase produces the narrow {\MgII}, {\SiIII}, and
{\SiIV} in a smaller cloud.
We obtained $\log U \simeq -2.6$ and the range $-1.0 \geq Z \geq -1.5$
for $0.0 \leq [\alpha / {\rm Fe}] \leq +0.5$. 
The higher ionization phase, with  $\log U \simeq -1.3$, has $b\sim
40$~{\kms}, which yields a size $S\sim 10$~kpc.
It is not possible for this high ionization component to be strictly
collisionally ionized nor is it possible for it to be a two--phase
photo plus collisionally ionized component; any contribution from a
collisionally ionized phase is insignificant.


\section{Discussion}
\label{sec:discussion}

As shown in Figure~\ref{fig:vpfits}, system A is comprised of six
distinct {\MgII} clouds with a total velocity spread of $\sim
200$~{\kms} and is $\sim -300$~{\kms} from system B, which has five
clouds spread over $\sim 100$~{\kms}. 
System C is $\sim +1000$~{\kms} from system B and is comprised of a
single, resolved {\MgII} cloud.
In Figure~3, we show the normalized FOS spectrum with
simulated spectra superimposed.
These low resolution data reveal that each system is rich in
multiple chemical species covering a wide range of ionization
potentials and that the chemical and ionization conditions differ from
system to system.

One motivation for our study was simply to ascertain if a multiphase
medium was required to explain the strong {\CIV}, {\NV} and {\OVI}
absorption lines.
In all three systems, we were required to postulate a higher
ionization component that is not seen in {\MgII} absorption.
We emphasize that (1) the overall {\MgII} cloud properties are well
constrained by the data and modeling, and (2) the allowed cloud to
cloud variations are constrained tightly enough that no scenario even
remotely modifies the requirement for a {\CIV}--{\NV}--{\OVI} high
ionization phase to explain the data.
To the accuracy afforded by the FOS spectrum, each of these high
ionization phases is well described by a single component with $b
\sim 70$~{\kms} (based upon simulations).
Given the large $b$ parameters and the range of sizes derived from the
models (Table~\ref{tab:DICs}), this high ionization gas is likely to
have a line of sight extent of $10 \leq S \leq 20$~kpc, and thus may
be a surrounding medium in which the $\sim 0.1$~kpc {\MgII} clouds are
embedded.

\subsection{Comparison of System Properties}

Consider the cloud to cloud variations in system B, which is a Lyman
limit system.  
The line of sight velocity spread of the {\MgII} clouds is $\sim
100$~{\kms}, which implies they are bound within a galactic potential.
If the clouds are equally illuminated by the extragalactic background,
then the presence of {\FeII} in three of the clouds (8, 9, and 11)
implies that they are more dense, more shielded from the ionizing
flux, and/or iron--group enriched relative to the other two clouds (7
and 10).
This suggests that clouds 8, 9, and 11 may be spatially contiguous
(relatively speaking), in that they may share similar histories of
iron--group enrichment from Type Ia SNe.
In the Galaxy, the association of Type Ia explosions with the
kinematically old disk implies that some events take place at large
scale heights, so there is uncertainty in how much iron--rich gas
is driven into galactic halos (\cite{wheeler89}).

The unique cloud in system B is cloud 10, which gives rise to a broader
absorption profile ($b  \simeq 13$~{\kms}) and has no detectable {\FeII}.
It has a higher ionization condition and gives rise to the majority of
the {\SiIV} absorption (see Figure~\ref{fig:stis}).
It also has the largest $N({\MgII})$.
Models yield that it is extended ($\sim 5$~kpc) and has lower
metallicity.
This leads us to conservatively speculate that cloud 10 is more akin
to a halo--like cloud or to a so--called Galactic high velocity cloud.
The ratio $N({\SiIV})/N({\MgII})$ may be a useful indicator
of the differing local environments of clouds in higher redshift
systems.

System C classifies as a ``weak'' {\MgII} absorbers, defined by
$W_{\rm r}(2796) < 0.3$~{\AA} (\cite{weakmgII}).
From a sample of thirty such systems over the redshift range $0.4 < z
< 1.4$, Churchill \etal found a wide range of
$W_{\rm r}({\FeII})/W_{\rm r}({\MgII})$ and 
$W_{\rm r}({\CIV})/W_{\rm r}({\MgII})$, presumedly due to variations in
abundance pattern and ionization conditions, including single phase
and multiphase.
These {\MgII} absorbers are sub--Lyman limit systems with $Z \geq -1$
and [some with $Z > 0$ (\cite{cl98})].
Apart from its line--of--sight proximity ($< 500$~{\kms}) to system B,
system A would classify as a weak {\MgII} absorber.
The $\sim 200$~{\kms} kinematic spread of the six 
clouds in system A, and the large $N({\CIV})/N({\SiIV})$ ratio in the
diffuse component are suggestive of lower ionization clouds moving
within a high ionization galactic corona, or halo, where the ratio
$N({\CIV})/N({\SiIV})$ is expected to be large (\cite{savage97}).
Is it possible that the system C {\MgII} cloud arises in a similar
environment as the six system A clouds, but that the line of sight
happens to sample only one cloud?
If so, the {\CIV} absorption strength in system C would be comparable
to that of system A, and it is significantly weaker.
For weak systems, it may be that strong, broad {\CIV} absorption implies
a larger number of {\MgII} clouds with a larger kinematic spread.

\subsection{Profile Anatomy: Model Predictions}
\label{sec:anatomy}

\begin{figure*}[th]
\figurenum{4}
\plotfiddle{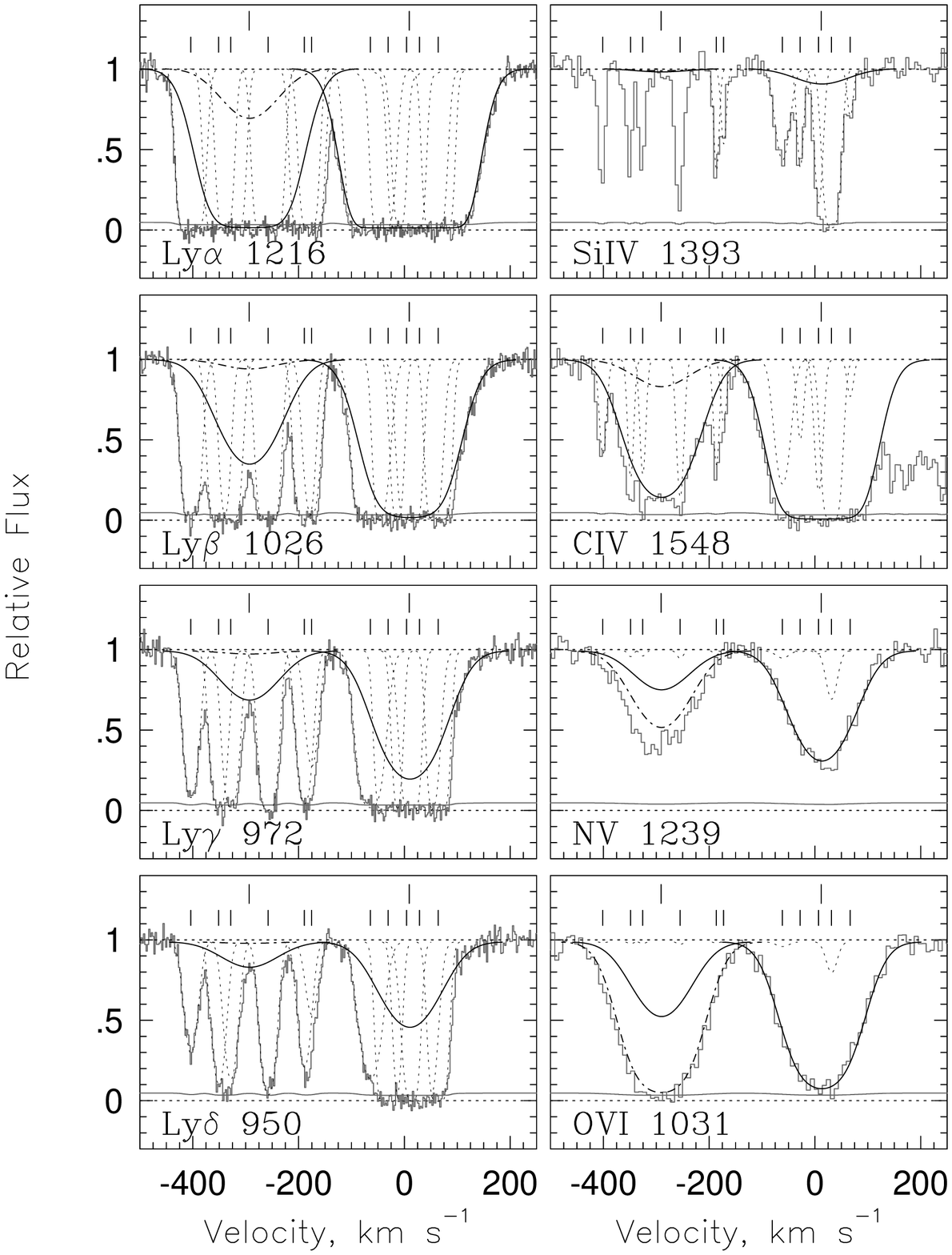}{6.5in}{0}{70}{70}{-227}{-37}
\protect\caption[Churchill.fig4.eps]
{\footnotesize
A selection of predicted high resolution profiles for systems A and B
based upon our models.  This simulated STIS/{\it HST\/} spectrum has
$R = 30,000$ ($v \sim 10$~{\kms}) and signal--to--noise ratio of 30.
Short ticks mark the velocities based upon the HIRES {\MgII} clouds
and long ticks mark the diffuse component centroids.  For system A,
the three--phase model (photoionized {\MgII} clouds and photo plus
collisionally ionized diffuse component) is shown and, for system B,
the two--phase model (photoionized clouds and diffuse component) is
shown.  The dotted--line curves are the profiles of the blended
individual {\MgII} clouds.  The solid--line curves are the
photoionized diffuse component profiles and the dash--dot curves are
of the collisionally ionized diffuse component.
\label{fig:stis}}
\end{figure*}

In Figure~\ref{fig:stis}, we present simulated STIS/{\it HST\/}
spectra with $R=30,000$, two pixels per resolution element, and a
signal--to--noise ratio of $30$ for the first four transitions in the
Lyman series, and for the {\SiIV}, {\CIV}, {\NV}, and {\OVI} profiles
of systems A and B.
These spectra were generated assuming Voigt profiles with the
properties listed in Tables~\ref{tab:PICs} and \ref{tab:DICs}.
These model profiles can be compared directly to observed data (from
STIS/{\it HST\/}) and thus provide a direct test of the models.
We show the contributing components as smooth curves.
The dotted--line curves are the {\MgII} clouds, with their velocity
centroids marked with the short ticks above the continuum.
The solid curves are from the photoionized diffuse component and
the dash--dot curves in system A represents the collisionally ionized
diffuse component.

\subsubsection{The Lyman Series}

Here, we emphasize the importance of the Lyman series.
For all three systems, the {\Lya} and {\Lyb} profiles in the FOS
spectra were significantly broader than could be fully accounted 
by the {\HI} obtained soley from the model {\MgII} clouds.
However, the narrower {\Lyg}, {\Lyd}, and {\Lye} profiles were
fully accounted by the {\HI} in these clouds.
Based upon the curve of growth behavior of the Lyman series, we found
that the addition of a very broad ($b>50$~{\kms}), lower $N({\HI})$
component naturally explained the deeper and broader {\Lya} and {\Lyb}
profiles, without overproducing the narrower {\Lyg}, {\Lyd}, and
{\Lye} profiles (or modifying the Lyman limit break).

Such behavior of the Lyman series in low resolution data is likely to
be a strong indication that a broad, perhaps highly ionized diffuse
component is present.
In Figure~\ref{fig:stis}, we illustrate this behavior as it would be
seen in higher resolution spectra.  
Note that the widths of {\Lya} and {\Lyb} are dominated by a
broad, high ionization component, whereas the higher order transition
widths are dominated by the {\MgII} clouds.
In high resolution spectra, a diffuse component gives rise to a
broadened, shallow wing when the profiles are saturated (see {\Lya} and
{\Lyb} in system B) and/or suppresses the recovery of the flux to the
continuum level between clouds in the profile centers (see {\Lyb} and
{\Lyg} in system A).

\subsubsection{High Resolution Metal Lines}

In system A, the {\CIV} profiles have structure due to the lower
ionization {\MgII} clouds.
The {\SiIV} absorption predominantly arises in the {\MgII} clouds, and
thus closely traces the {\MgII} kinematics.
Together, these profiles are tantalizing similar to those observed by
Savage, Sembach, \& Cardelli (1994\nocite{savage94}) along the line of
sight to HD~167756 in the Galaxy. 
We quote, ``The sight line contains at least two types of highly
ionized gas.  One type gives rise to a broad {\NV} profile, and the
other results in a more structured {\SiIV} profile.  The {\CIV}
profile contains contributions from both types of highly ionized
gas.''
We also find similarities between the model {\CIV}
profiles and those in the damped {\Lya} systems at $z_{\rm
abs}=3.3901$ toward Q$0000-262$ and $z_{\rm abs}=2.2931$ toward
Q$0216+080$ (Lu \etal 1996\nocite{lu96}, Figures 2 and 3).

In system B, cloud 10 is clearly unique among the {\MgII} clouds.
This cloud has the largest ionization parameter and accounts for the
majority of the {\SiIV} absorption, which was clearly constrained to
arise in the {\MgII} clouds.
Since the ionization conditions of clouds 8, 9, and 11 were set by
their measured $N({\FeII})/N({\MgII})$, the ionization condition in
cloud 10 was well constrained by the ratio $N({\SiII})/N({\SiIV})$;
there was no alternative but for cloud 10 to dominate the {\SiIV}
absorption.
The point is that this cloud may arise in a spatially distinct
environment and have a unique formation history from the other clouds
in system B (see \S~\ref{sec:discussion}).
The {\CIV} profile is dominated by a diffuse higher
ionization phase, whereas the {\SiIV} is predominantly due to higher
density clouds.
Some lines of sight through the Galaxy also exhibit narrow {\SiIV}
profiles and broad {\CIV} profiles (for examples, see
\cite{sembach94}; \cite{savsem94}).
In contrast, the {\SiIV} and {\CIV} profiles in the $z_{\rm
abs}=2.8268$ damped {\Lya} system towards Q$1425+603$ (\cite{lu96})
appear to arise in the same phase.

A synthetic STIS/{\it HST\/} spectrum for System C (not shown) reveals
a narrow {\SiIV} profile and broad {\CIV}, {\NV}, and {\OVI} profiles.
The {\CIV} profile exhibits a slightly deep, narrow core due to the
low ionization phase.
However this contribution would be lost for the expected noise levels
in observed data, unless this narrow component was off--center
by $\sim 40$~{\kms} relative to the broad component (within
the uncertainty of our modeling such an offset is not ruled out).

\section{On the Nature of the High Ionization Phase}
\label{sec:onthe}

Within 100{\arcsec} of the QSO, Kirhakos \etal
(1994\nocite{kirhakos94}) identified $10$ galaxies with $21 \leq g
\leq 22$.
Compared to field galaxies, this is a slight overdensity by a factor
of a few to several (\cite{tyson88}).
Identified within 10{\arcsec} of the QSO are three bright galaxies
with impact parameters $29$, $45$, and $47~h^{-1}$~kpc, respectively
($q_0 = 0.05$). 
At $z = 0.93$, Thimm (1995\nocite{thimm95}) detected a {\OII} $\lambda
3727$ flux of $9 \times 10^{-17}$ ergs {\cmsq} s$^{-1}$ from the
galaxy at $45~h^{-1}$~kpc.

Given these facts, we entertain the possibility that the {\MgII}
absorption systems arise in individual galaxies either in a small
group or possibly in a cluster.  
Such a possibility would have interesting implications in view of the 
types of mechanisms that could give rise to highly ionized gas in such
environments.

\subsection{Intragroup Corona?}

Mulchaey \etal (1996\nocite{mulchaey96}) predicted that poor groups
that are rich in spiral galaxies have hot, diffuse coronae that are
cooler than the X--ray coronae surrounding poor groups dominated by
E/S0 galaxies.
This high ionization intragroup material is predicted to have $T\sim 2
\times 10^{6}$~K and to primarily give rise to strong {\OVI}, whereas
any {\CIV} or {\NV} absorption is predicted to arise in the proximity
of the galaxies themselves (\cite{mulchaey96}). 
As such, the {\OVI} profiles would be broader than the {\NV} and
{\CIV}.
The same would hold for the Lyman series (\cite{verner94}).

Our models can fully explain the {\OVI}, {\NV}, and {\CIV} in a single
component of $T \sim 3 \times 10^{4}$~K gas (the possible collisional
component in system A has $T \sim 3 \times 10^{5}$~K).
The observed {\OVI} or {\HI} widths are not broader than those of the
{\CIV} and {\NV}, and their $\sim 70$~{\kms} $b$ parameters are not
suggestive of gas that is kinematically akin to the $\sim 200$~{\kms}
dispersion of a galaxy group.
Furthermore, the {\OVI}, {\NV}, and {\CIV} profiles are clearly
aligned with the three {\MgII} absorption redshifts, suggesting that
the high ionization material is spatially coincident with the
individual {\MgII} systems.
The upper limit on the size of the photoionized diffuse components
is $S\leq30$~kpc (assuming photoionization equilibrium), and the
inferred $Z\sim 0$ metallicities suggest that the high ionization
material surrounds the galaxies and has been enriched by them. 
Significantly lower metallicities are expected if the gas was
intragroup material left over from the formation processes.
We do not favor an interpretation of the data in which the
high ionization material arises in a intragroup or common halo.

\subsection{Galactic Coronae?}

The highly ionized material in these three $z\sim 1$ galaxies may
be more akin to galactic coronae (see Spitzer
1956\nocite{spitzer56}, 1990\nocite{spitzer90}; Savage \etal
1997\nocite{savage97}), material stirred up by energetic mechanical
processes, such as galactic fountains.
In this scenario, the gas is concentrated around individual galaxies
which presumably provide a source of support, heating, and chemical
enrichment.  

In the Galaxy, $b \simeq 60$~{\kms} was measured for {\CIV}, and was 
observed to slightly increase with the ionization level of the
transition (\cite{savage97}).  
Simulations of the allowed range of $b$ parameters for systems A and B
yielded similar Doppler widths, with $b \simeq 70$~{\kms} providing
a good match to the {\CIV}, {\NV}, and {\OVI} profiles for both
systems.
If the high ionization diffuse components in these systems are
arising in coronae analogous to that surrounding the Galaxy, they may
have similar turbulent processes supporting their scale heights.

The radial extent of the Galactic corona is unknown, but for local
galaxies, radio maps reveal {\HI} extends to tens of kpc
(\cite{corbelli89}; \cite{vangorkom93}), beyond which the hydrogen
becomes optically thin and a highly ionized extension is expected
(\cite{maloney93}; \cite{corbelli93}; \cite{dove94}).
The impact parameters of the candidate galaxies are $\sim
30$--$50~h^{-1}$~kpc, with the orientation of the line of sight through
each galaxy unknown.
Absorption, even from non--spherical absorbers, is not unexpected at
an impact parameter of $\sim 50$ kpc (see Figure~1 of \cite{cc96}).
For the Galaxy, Savage \etal (1997\nocite{savage97}) measured the
effective scale heights of $5.1$, $4.4$, and $3.9$~kpc for {\SiIV},
{\CIV}, and {\NV}, respectively.
We find a preferred size of $10$--$20$~kpc for the size of the high
ionization gas at $z\sim 0.93$, with an upper limit of $\sim 30$~kpc.
Since these ``sizes'' (or the path length through an oriented
structure) are comparable to twice the Galactic scale height, the
distribution of high ionization is consistent with Galactic--like
coronae. 

A scenario in which the high ionization gas arises in individual
galactic coronae is in contrast to that proposed by Lopez \etal
(1998\nocite{lopez98}) for a system at $z\sim 1.7$ in the spectra of
HE~$1104-181$~A, B.
In a double line of sight study, Lopez \etal found {\OVI} profiles
consistent with $110 \leq b \leq 180$~{\kms} and that the extent of
the highly ionized gas was $\sim 100$~kpc. 
These inferences also differ from those of Bergeron \etal
(1994\nocite{bergeron94}), who found that the {\OVI} phase in the
$z_{\rm abs} \sim 0.8$ {\MgII} absorber toward PKS~$2145+064$ was at
least $50$~kpc in extent.


\section{Conclusions}
\label{sec:conclusion}

We have studied the kinematic, chemical, and ionization conditions of
three metal--line absorption systems at $z\sim 1$ seen in the
PG~$1206+459$ spectrum.
The  systems were selected by the presence of {\MgII} absorption in a
high resolution spectrum, and were chosen as a pilot study of a larger
program designed to chart the physical conditions and evolution of
absorbing gas in galaxies.
The {\MgII} profiles are shown in Figure~\ref{fig:order28}, with each
system designated A, B, and C.
Rich absorption line data from FOS/{\it HST\/} (\cite{kp13}) revealed
strong {\CIV}, {\NV}, and {\OVI} absorption, as well as {\HI} and many
other species and transitions covering a wide range of ionization
potentials.
Ground--based imaging data (\cite{kirhakos94}) revealed three
candidate absorbing galaxies within 10{\arcsec} of the QSO and
possibly a group of galaxies within 100{\arcsec} of the QSO.

The main goals of the study were to see if multiphase gas was
required to explain the strong high ionization absorption line data,
and to infer some level of information on the gas metallicities and
spatial distributions.
Assuming the {\MgII} clouds are photoionized, we found the range of
chemical and ionization conditions consistent with both the high
resolution and low resolution data.
We then postulated the presence of high ionization components not seen
in {\MgII} absorption to explain the unaccounted {\CIII}, {\CIV},
{\NV}, and {\OVI} absorption.

We briefly summarize the main results of our study:

1. 
For systems A and B, we were required to postulate a high ionization
phase in addition to the lower ionization {\MgII} clouds.
System C could be made marginally consistent with a single--phase
absorber, though a two--phase absorber is strongly preferred due to
the nature of the Lyman series absorption.
We infer that, in these systems, the lower ionization {\MgII} clouds
arise in high ionization diffuse gas; each of these absorption systems is
comprised of a multiphase gaseous medium.
We find that the high ionization phase of system A {\it could\/} arise
in multiple, narrower  ``{\CIV} clouds''.
For system B, such {\CIV} clouds are ruled out.

2.
The absorbing gas in both the {\MgII} clouds and the high ionization
components are consistent with photoionized clouds.  
In systems B and C, a collisionally ionized phase is ruled out.
Only in system A could the data be made consistent with a three--phase
absorber, incorporating the photoionized {\MgII} clouds, a highly
photoionized diffuse component, and a collisionally ionized component.
This three--phase model provided a more consistent match to the {\NV}
absorption in this system.
In this three--phase scenario, the {\CIV} could arise in a few
narrower components.
However, $[{\rm N}/{\rm O}] \simeq 0.15$ in the highly photoionized
component, instead of the assumed $[{\rm N}/{\rm O}] = 0$, would
remove the need for the collisionally ionized gas and rule out the
three--phase absorber.

3. 
We find no evidence that the {\OVI} gas is in a separate
and very highly ionized diffuse phase that encompasses the {\CIV} and
{\NV} absorption.
Based upon the $b\sim 70$~{\kms} profile widths, inferred $Z\sim 0$
metallicities, $3\times 10^{4}$~K temperatures, inferred $S\leq
30$~kpc sizes, and the clear redshift alignment of the high ionization
transitions with the {\MgII} systems, we suggest that the high
ionization gas is analogous to the Galactic corona in that it traces
the galaxies themselves and does not appear to exhibit the
characteristics predicted for intragroup or intracluster material. 
The {\OVI} likely arises in the same phase as the {\CIV} and {\NV}.
The {\SiIV} is constrained to arise in the same phase as the {\MgII}
clouds, whereas {\CIII} arises in both the clouds and the high
ionization phase.

4.
We have found cloud to cloud variations in the chemical and ionization
conditions in the five {\MgII} clouds of system B.
Three of the clouds are likely to be iron--group enriched and have
higher densities and low ionization conditions.
The majority of the neutral hydrogen giving rise to the Lyman break in
the FOS spectrum is from these clouds.
A lower metallicity {\MgII} cloud giving rise to a broader absorption
profile is interspersed in velocity with these clouds, likely has an
$\alpha$--group enhanced abundance pattern, and gives rise to the
majority of the {\SiIV} absorption.
We speculate that this cloud may be similar to a halo--like cloud,
and suggest that the ratio $N({\SiIV})/({\MgII})$ may be a useful
indicator for discriminating between clouds in different parts of 
high redshift galaxies.

The most compelling reason why we favor the scenario in which the high
ionization diffuse material is coupled to the galaxies is the clear
kinematic separation of the {\OVI} profiles corresponding to systems A
and B.
Each of the inferred diffuse components must be centered (at least
roughly) on the systems, and they must be distinct from one another
in velocity space.  
There are additional, if less compelling, arguments.
The inferred $b$ parameters in the model diffuse components are
in the same regime as those found for the Galactic corona
(\cite{savage97}), further suggesting that the material is galaxy
associated.
The inferred metallicities are high, $Z \sim 0$, which is best
understood if the material had been enriched by its host galaxy and
further suggests that the origin and source of enrichment of the
diffuse gas is related to star forming parts of galaxies.
The enrichment could be due to {\it in situ\/} star formation as gas
clouds collide and cool in the galactic halos (\cite{ss92}),
or could be due to galactic fountain processes from the galactic disks
(\cite{spitzer90}).

As a speculative aside, we ask if Galactic--like coronae {\OVI} 
absorbers are likely to be a common form of {\OVI} systems at
$z\leq 1$, as opposed to group--halos.
Burles \& Tytler (1996\nocite{bt96}) have shown that absorbers
selected by the presence of {\OVI} have the same redshift path density
as Lyman limit--{\MgII} absorbers at $\left< z \right> = 0.9$.
As we have found here, {\OVI} can be associated with a Lyman limit
system when the absorbing gas is segregated into multiple ionization
phases.
If multiphase absorption is common in {\MgII} absorbers, some {\OVI}
might arise in Galactic--like corona.

It would seem that this pilot study has shown that wholesale study of
the kinematic, chemical, and ionization conditions of {\MgII}
absorbers, using high resolution {\MgII} profiles and the available
low resolution {\it HST} spectra, would yield a improved understanding
of galactic gas at early epochs.
In order to assess the robustness of our modeling, we have synthesized
high resolution STIS/{\it HST\/} spectra of the ultraviolet
transitions (presented in Figure~\ref{fig:stis}).
The modeling techniques applied in this paper can be directly tested
by comparing these predicted profiles with those observed with
STIS/{\it HST}.
If our approach proves to yield an accurate description of the gas,
then wholesale modeling can be embarked upon for roughly 50 {\MgII}
systems without requiring large amounts of space based telescope time
to acquire high resolution spectra of high quality.


\acknowledgments

This work was supported by NSF AST--9529242 and AST--9617185 and by NASA
NAG5--6399 and AR--07983.01--96A from STScI.  
Special thanks to Buell Jannuzi, Sofia Kirhakos, and Don Schneider for
generously supplying the FOS/{\it HST\/} spectra and for helpful
discussion regarding the reduction and analysis of the FOS data.
We are thankful for comments from the anonymous referee that 
resulted in an improved manuscript.
We thank Karen Knierman and Jane Rigby for assistance in producing the
stacked spectrum of the system A clouds for the limit on $N({\FeII})$,
and are grateful to Steven S. Vogt for the HIRES spectrograph.


\appendix

\section{Application of Observed Constraints}
\label{app:constraints}

\begin{figure*}[b]
\figurenum{A1}
\plotfiddle{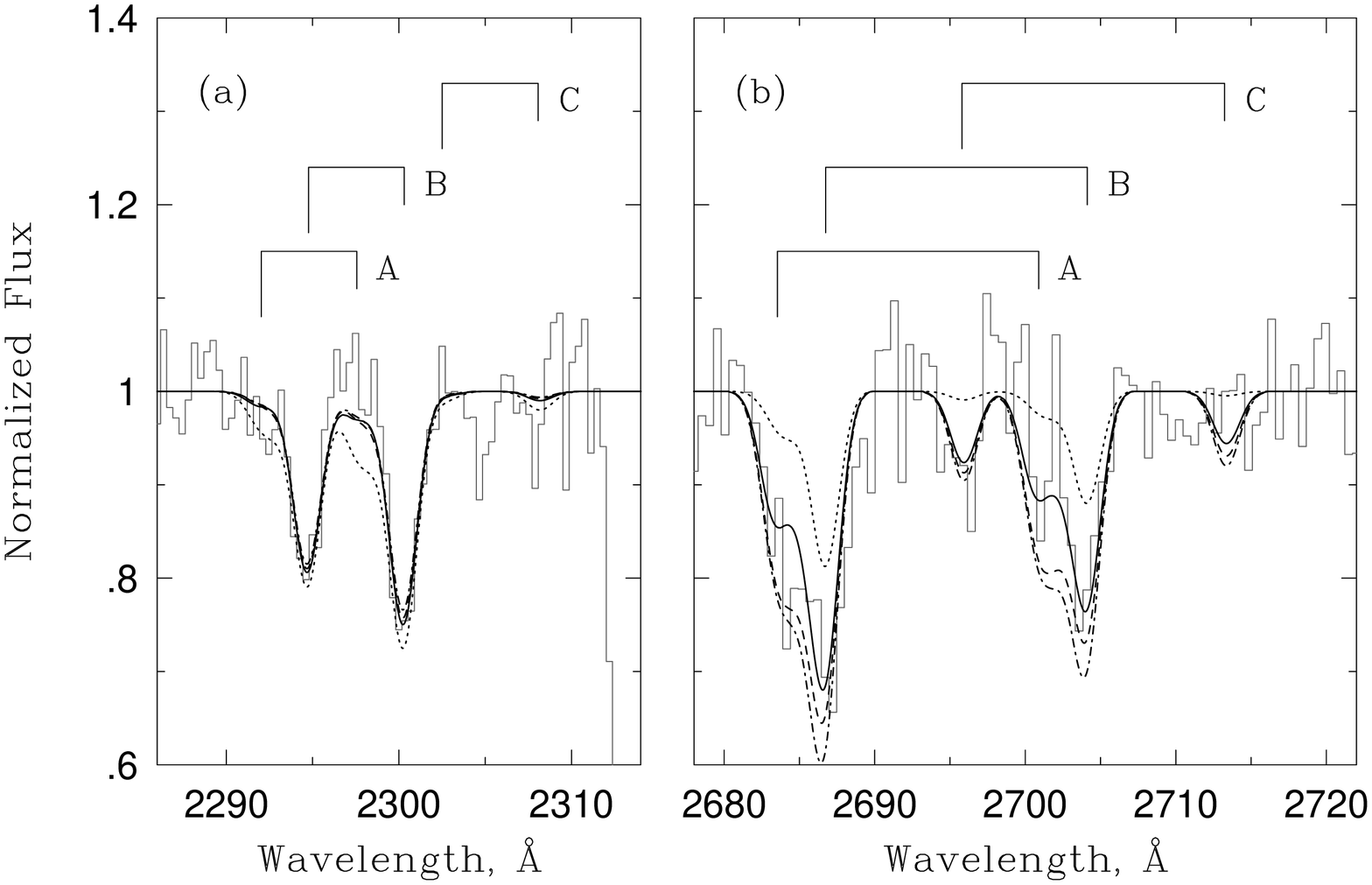}{3.3in}{0}{60}{60}{-236}{-50}
\protect\caption
{\footnotesize
An illustration of how the {\SiIIdblt} (panel $a$) and {\SiIVdblt}
(panel $b$) doublets were used to constrain the ionization conditions
for the photoionized {\MgII} clouds in systems A, B, and C.  
The clouds are tuned by the ratio $f=\log \{ N({\FeII})/N({\MgII}) \}$,
which yields a unique ionization parameter, $U$ (clouds with measured
{\FeII} are fixed).  The dotted--line curve is for the $1\sigma$ limits
on $f$ for each cloud as measured from the HIRES data.  The solid,
dash, and dot--dash curves are for $f = -3.0$, $-3.5$, and $-4.0$,
respectively.  The corresponding ionization parameters are $\log
U=-2.30$, $-2.15$, and $-2.00$ (a small range, indeed).  The ratio
$N({\SiII})/N({\SiIV})$ decreases with decreasing $f$, or
equivalently, as the ionization parameter is increased.
\label{fig:SiII-SiIV}}
\end{figure*}

In this appendix, we demonstrate our modeling methodology for
constraining the ionization parameters, metallicities, and overall
{\MgII} cloud properties using the FOS/{\it HST\/} data.

In Figure~\ref{fig:SiII-SiIV}, we present an example of how the data
are used to constrain the ionization parameter, $U$.
In Figure~\ref{fig:SiII-SiIV}$a$ the synthetic FOS spectrum of the
{\SiIIdblt} doublet is superimposed on the FOS data for the $1\sigma$
upper limits on $\log \{ N({\FeII})/N({\MgII}) \}$, and for the set
ratios $-3.0$, $-3.5$, and $-4.0$.
The {\SiIVdblt} doublet is shown in Figure~\ref{fig:SiII-SiIV}$b$.
The assumed abundance pattern is solar.
Clouds 8, 9, and 11 have their $U$ constrained by the observed
$\log \{ N({\FeII})/N({\MgII})\} $.  
The ratio of {\SiII} to {\SiIV} can uniquely determine $U$ for the
remaining clouds.
For System A, the best match is provided by $\log U = -2.3$ [which
corresponds to $\log \{ N({\FeII})/N({\MgII}) \}  = -3.0$].
Lower ionization clouds are not possible because they overproduce
{\SiII}, whereas higher ionization clouds overproduce {\SiIV}.  
For System B, a reasonable match to the observed {\SiII} and {\SiIV}
is achieved if clouds 7 and 10 (dominated by 11 because of its larger
{\MgII} column density) are assigned $\log U = -2.0$, which
corresponds to $\log \{ N({\FeII})/N({\MgII}) \}  = -4.0$.
Less ionized clouds are possible, but an additional more highly 
ionized component would then be required to produce {\SiIV}.

\begin{figure*}[hb]
\figurenum{A2}
\plotfiddle{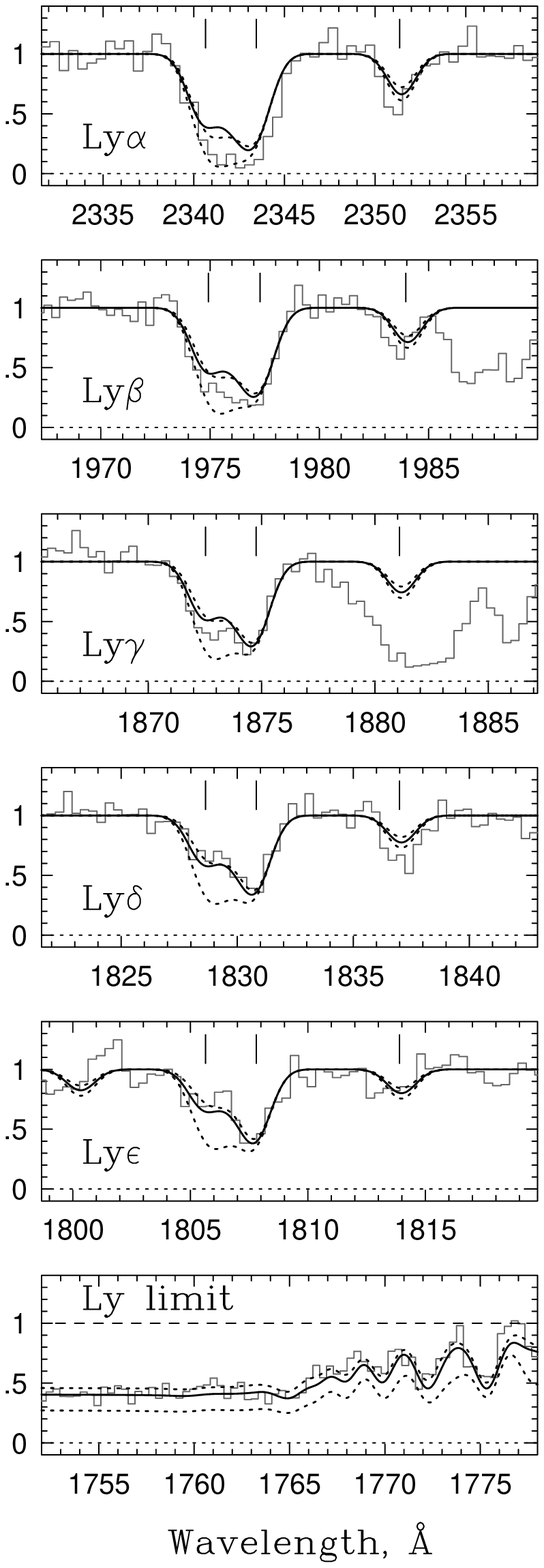}{6.3in}{0}{70}{70}{-152}{-42}
\protect\caption[Churchill.figA2.eps]
{\footnotesize
An illustration of how the Lyman series was used for tuning the
metallicity, $Z$, and neutral hydrogen column density, $N({\HI})$, for
set ionization conditions.  Only the photoionized {\MgII} clouds are
shown. For these clouds the ionization conditions were tightly
constrained by the {\SiII}, {\SiIII}, and {\SiIV} data.  The ticks
mark systems A, B, and C, from left to right.  A solar abundance
pattern is assumed.  The thick solid model has $Z=0$ (solar); the
dotted--line model with stronger $N({\HI})$ absorption has $Z=-0.4$;
and, the dotted--line  model with the weaker absorption has $Z=+0.4$.
Note that the series is not fit self--consistently for any value
of $Z$ (see \S~\ref{sec:anatomy}). \label{fig:HI-Z}}
\end{figure*}

In Figure~\ref{fig:HI-Z}, we present an example of how the Lyman series
and limit are used to constrain the metallicity for the ionization
parameters determined in the above illustration.
We have assumed $\hbox{[$\alpha$/Fe]} = 0$ for this illustration.
Three metallicities are shown, $Z = -0.4$, $0.0$, and $+0.4$, to
illustrate the strength variations in the Lyman series and limit.
At low metallicity, a large $N({\HI})$ is needed to produce the
observed metal lines with low and intermediate ionization levels.
Such a large $N({\HI})$ can be inconsistent with the observed Lyman
series lines and Lyman limit break.  
As described in \S~\ref{sec:data}, the Lyman limit break implies a total
$N({\HI})$ of $10^{17.2}$~{\cmsq}. 
For System A, the best match to the Lyman series lines is given by $Z =
+0.2$, super--solar metallicity.
For System B, the best match also has high metallicity;
clouds 8, 9, and 11 must have $Z = 0.0$, and cloud 10 must also have
near solar metallicity in order that there is not too large an
$N({\HI})$.
It is important to point out that the assumed abundance pattern directly
affects the inferred metallicities.
If the abundance pattern is $\alpha$--group enhanced,
$\hbox{[$\alpha$/Fe]} = +0.5$, the illustrated metallicities would
proportionally drop by $\sim 0.5$ dex.
This is because the cloud properties are tuned to the $N({\MgII})$,
and magnesium is an $\alpha$--group element (see \S~\ref{sec:models}).

\section{Galactic Ionizing Photons?}
\label{app:models}

Given the {\OII} $\lambda 3727$ emission measured by Thimm
(1995\nocite{thimm95}), it is reasonable to assume that high energy
photons could be escaping the galaxies (e.g.\ \cite{bergeron94}).
The Thimm measurement of an {\OII} flux from one
galaxy (and limits for the others) constrains
the contribution of galactic ionizing photons to be
$< 7 \times 10^5$~{\cmsq}~s$^{-1}$ under the assumption that 50\% of
the photons escape from the galaxy.
This is a factor of five to 50 times greater than that estimated for
the Galaxy (\cite{bland-hawthorn}) and for external galaxies
(\cite{jean-michel}), respectively. 
Keeping this generous upper limit in mind, we consider the effect
a change in spectral shape could have on our main results.
Several contrasting starburst models from Bruzual \& Charlot
(1993\nocite{bruzual}) were explored.
In all cases, the galaxy spectra were normalized relative to the
Haardt \& Madau (1996\nocite{handm96}) flux, $F_{HM}$, at 1 Rydberg.  
Depending on the spectral shape, the above limit corresponds to
$F_{sb}/F_{HM} < 4$--$10$, where $F_{sb}$ is the flux of the starburst
galaxy model at 1 Rydberg.

First we consider a $10^7$ year instantaneous burst model,
because of its prominent {\HeI} edge [see Fig.~4 of Bruzual
\& Charlot (1993\nocite{bruzual})].  
As the ratio $F_{sb}/F_{HM}$ is increased, the hydrogen becomes more
ionized and it is possible to decrease the cloud metallicities without
overproducing {\HI}.  
However, only a modest decrease of 0.4 dex is possible for
$F_{sb}/F_{HM} \leq 10$.  
For these models, there are relatively fewer photons capable of
ionizing {\CIII} and thus there is even less {\CIV} produced in the
{\MgII} clouds.  
Similarly, in order to maintain the observed {\SiII} to {\SiIV} ratio,
it is necessary to slightly increase the cloud ionization parameter by
a few tenths of a dex from the pure Haardt \& Madau case.  
We also consider the effect of the change in spectral shape upon the
diffuse phase that produces {\CIV} and {\OVI}.  
Similar conclusions hold for the metallicity, which can decrease by at
most 0.5 dex from the pure Haardt \& Madau case; the diffuse phase still
must have relatively high metallicity ($> 0.3Z$).  
For energies greater than the {\HeII} edge, this overall spectrum is
unchanged relative to Haardt \& Madau, and therefore the balance
between {\CIV} and {\OVI}, and other high ionization energy
transitions is unaltered.

In order to explore a severe {\HI} edge, we explored another extreme
model, a $10^8$ year instantaneous burst.
In this case the measured {\OII} constraint also forces the upper limit
$F_{sb}/F_{HM} \leq 10$.  
Interestingly, the {\MgII} cloud metallicity cannot be reduced
(relative to pure Haardt \& Madau) because $N({\HI})$ actually
increases  relative to $N({\MgII})$ as the galaxy contribution is
increased.
Due to similar behavior, the metallicity of the required diffuse
component is constrained to be solar or super--solar.
The change in spectral shape has the effect of increasing {\SiIV} and
{\SVI}, but decreasing {\CIII}.
The bulk of the {\SiIV} still arises in the {\MgII} clouds, but {\SVI}
in the diffuse phase is overproduced, and the required super--solar
metallicity seems implausible.

Finally, we consider a later--type galaxy model with a 16~Gyr stellar
population and with exponentially decreasing but persistent, star
formation [the $\mu=0.01$ model (\cite{bruzual83})].
This model has 1\% of the total star--forming mass in stars after 1
Gyr.
The star formation longevity of this relatively quiescent galaxy is
more plausible than an instantaneous burst.
As the galactic flux is incrementally increased relative to the
extragalactic background, the metallicity constraints are unchanged,
but the ratio $N({\CIV})/N({\MgII})$ decreases due to the drop in flux
at the {\HeII} edge.
Thus, the requirement for a diffuse component in these absorbers is
strengthened if galactic flux is contributing.
We also found that it is increasingly difficult to produce
the required $N({\OVI})/N({\CIV})$ ratio as the galactic flux is
increased unless the ionization parameter is significantly increased.
This results in the destruction of {\SiIV}.
Thus, our conclusion that the {\SiIV} must arise in the {\MgII}
clouds is also strengthened if such a galactic flux is contributing to
or dominating the ionizing spectrum.
Likewise, our conclusion holds that any collisionally ionized diffuse
component in system B must be negligible.
As with the extragalactic background scenario, the {\CIII} becomes too
large and inconsistent with the data if one lowers the ionization
condition in the photoionized diffuse component.

Although some starburst galaxy spectra can somewhat decrease the
constraint on the cloud and diffuse component metallicities, this
effect is limited to less than 0.5 dex due to constraints on {\OII}
emission.
These galaxies are not strong starbursts, though they could have
modest galaxy contribution to the spectral shape.
Thus, we find that our {\it general\/} conclusions with regard to the
requirement of highly ionized diffuse components and relatively high
metallicity clouds are not sensitive to the assumed spectral shape of
the ionizing flux, though the details of the adopted models would be
somewhat modified.  



\newpage


\begin{deluxetable}{crcrccccccr}
\tablenum{1}
\tablecolumns{11}
\tablewidth{0pc}
\tablecaption{HIRES/Keck {\MgII} and {\FeII} Data}
\tablehead
{
&  &  &  & &
\multicolumn{2}{c}{\MgII} & & \multicolumn{2}{c}{\FeII} & 
 \\ 
\cline{6-7} \cline{9-10} 
\colhead{Sys} & \colhead{Cld} & \colhead{$z_{\rm abs}$} & \colhead{$v$} & &
\colhead{$\log N$} &
\colhead{$b$} & &
\colhead{$\log N$\tablenotemark{a}} &
\colhead{$b$} &
\colhead{{\FeII}/{\MgII}} \\
& No. &  &   
\colhead{[{\kms}]} & &
\colhead{[{\cmsq}]} &
\colhead{[{\kms}]} & &
\colhead{[{\cmsq}]} &
\colhead{[{\kms}]} & 
\colhead{[$\log$]} }
\startdata
A &  1 & 0.92501 & $-403.5$ & & $11.92\pm0.03$ & $ 5.98\pm0.55$ & & $<11.47$       & $\cdots$       & $<-0.5$  
\\
A &  2 & 0.92535 & $-350.7$ & & $12.16\pm0.02$ & $ 2.89\pm0.26$ & & $<11.48$       & $\cdots$       & $<-0.7$  
\\
A &  3 & 0.92550 & $-327.7$ & & $12.12\pm0.02$ & $ 2.98\pm0.28$ & & $<11.48$       & $\cdots$       & $<-0.6$  
\\
A &  4 & 0.92595 & $-256.8$ & & $12.35\pm0.01$ & $ 4.77\pm0.20$ & & $<11.48$       & $\cdots$       & $<-0.9$  
\\
A &  5 & 0.92639 & $-188.0$ & & $12.01\pm0.02$ & $ 3.10\pm0.37$ & & $<11.49$       & $\cdots$       & $<-0.5$  
\\
A &  6 & 0.92648 & $-174.2$ & & $11.60\pm0.05$ & $ 3.69\pm0.89$ & & $<11.49$       & $\cdots$       & $<-0.1$  
\\
B &  7 & 0.92720 & $ -62.8$ & & $11.72\pm0.07$ & $16.19\pm3.54$ & & $<11.47$       & $\cdots$       & $<-0.3$  
\\
B &  8 & 0.92742 & $ -29.0$ & & $13.44\pm0.04$ & $ 5.67\pm0.15$ & & $12.75\pm0.02$ & $ 6.19\pm0.27$ & $ -0.7$  
\\
B &  9 & 0.92764 & $   6.0$ & & $13.29\pm0.01$ & $ 7.38\pm0.14$ & & $12.21\pm0.06$ & $ 8.52\pm1.26$ & $ -1.1$  
\\
B & 10 & 0.92780 & $  30.4$ & & $12.60\pm0.01$ & $12.70\pm0.49$ & & $<11.45$       & $\cdots$       & $<-1.2$  
\\
B & 11 & 0.92803 & $  66.0$ & & $12.82\pm0.01$ & $ 5.08\pm0.10$ & & $12.07\pm0.05$ & $ 2.85\pm0.76$ & $ -0.8$  
\\
C & 12 & 0.93428 & $1038.3$ & & $12.05\pm0.02$ & $ 7.52\pm0.52$ & & $<11.41$       & $\cdots$       & $<-0.6$  
\\
\enddata
\tablenotetext{a}{The mean upper limit on $\log N({\FeII})$ is
11.1~{\cmsq} for the six clouds in system A based upon the technique of
stacking the {\MgII} clouds.}
\label{tab:vpfits}
\end{deluxetable}

\begin{deluxetable}{rrccrccl}
\tablenum{2}
\tablewidth{386pt}
\tablecolumns{8}
\tablecaption{FOS/HST Identified/Constraint Lines}
\tablehead
{
\colhead{No.} &
\colhead{$\lambda$} &
\multicolumn{2}{c}{Line ID} &
\colhead{IP} &
\colhead{Sys} & &
\colhead{Notes}  \\ 
\cline{3-4}
 & \colhead{[{\AA}]} & 
\colhead{Ion} & 
\colhead{$\lambda$, {\AA}} &
\colhead{[eV]} & & }
\startdata
   1a &  1762.92 &    {\NII} &  915.61 & 29.6  &  A &   &     \\ 
   2a &  1765.03 &    {\NII} &  915.61 &       &  B &   & matches depression at Lyman limit    \\ 
   3a &  1770.98 &    {\NII} &  915.61 &       &  C &   &     \\ 
   4a &  1797.13 &    {\SVI} &  933.38 & 88.0  &  A &   & C-fit?    \\ 
   5a &  1799.19 &    {\SVI} &  933.38 &       &  B &   & Bl--{\Lysix} from sys C; C--fit?; Note\tablenotemark{a}  \\ 
   6a &  1805.42 &    {\SVI} &  933.38 &       &  C &   & Bl--7a; C-fit? \\ 
   7a &  1805.65 &    {\Lye} &  937.80 & 13.6  &  A &   & Bl--6a; C-fit? \\ 
   8a &  1807.80 &    {\Lye} &  937.80 &       &  B &   &     \\ 
   9a &  1813.90 &    {\Lye} &  937.80 &       &  C &   &     \\ 
  10a &  1818.58 &    {\SVI} &  944.52 & 88.0  &  A &   &     \\ 
  11a &  1820.66 &    {\SVI} &  944.52 &       &  B &   &     \\ 
  12a &  1826.97 &    {\SVI} &  944.52 &       &  C &   & Bl--13a \\ 
  13a &  1828.64 &    {\Lyd} &  949.74 & 13.6  &  A &   & Bl--12a \\ 
  14a &  1830.82 &    {\Lyd} &  949.74 &       &  B &   &     \\ 
  15a &  1836.99 &    {\Lyd} &  949.74 &       &  C &   &     \\ 
  16b &  1872.52 &    {\Lyg} &  972.54 &       &  A &   &     \\ 
  17b &  1874.76 &    {\Lyg} &  972.54 &       &  B &   &     \\ 
  18b &  1881.08 &    {\Lyg} &  972.54 &       &  C &   & Bl--19b; Bl--{\Lya} \\ 
  19b &  1881.15 &   {\CIII} &  977.02 & 47.9  &  A &   & Bl--18b \\ 
  20b &  1883.40 &   {\CIII} &  977.02 &       &  B &   &     \\ 
  21b &  1889.75 &   {\CIII} &  977.02 &       &  C &   &     \\ 
  22b &  1905.76 &   {\NIII} &  989.80 & 47.4  &  A &   & Bl--{\Lya} or {\Lyb}?    \\ 
  23b &  1908.04 &   {\NIII} &  989.80 &       &  B &   &     \\ 
  24b &  1914.47 &   {\NIII} &  989.80 &       &  C &   &     \\ 
  25c &  1974.93 &    {\Lyb} & 1025.72 & 13.6  &  A &   &     \\ 
  26c &  1977.28 &    {\Lyb} & 1025.72 &       &  B &   &     \\ 
  27c &  1983.95 &    {\Lyb} & 1025.72 &       &  C &   &     \\ 
  28c &  1986.87 &    {\OVI} & 1031.93 & 138.1 &  A &   &     \\ 
  29c &  1989.25 &    {\OVI} & 1031.93 &       &  B &   &     \\ 
  30c &  1995.36 &    {\CII} & 1036.33 &  24.3 &  A &   & Bl--31c; Bl--{\Lya}? \\ 
  31c &  1995.95 &    {\OVI} & 1031.93 & 138.1 &  C &   & Bl--30c; Bl--{\Lya}? \\ 
  32c &  1997.75 &    {\CII} & 1036.34 &  24.3 &  B &   & Bl--33c \\ 
  33c &  1997.83 &    {\OVI} & 1037.62 & 138.1 &  A &   & Bl--32c \\ 
  34c &  2000.21 &    {\OVI} & 1037.62 &       &  B &   & Bl--{\Lya}?    \\ 
  35c &  2004.48 &    {\CII} & 1036.34 &  24.3 &  C &   &     \\ 
  36c &  2006.96 &    {\OVI} & 1037.62 & 138.1 &  C &   &     \\ 
  37e &  2292.03 &   {\SiII} & 1190.42 &  16.3 &  A &   &     \\ 
  38e &  2294.76 &   {\SiII} & 1190.42 &       &  B &   &     \\ 
  39e &  2297.56 &   {\SiII} & 1193.29 &       &  A &   &     \\ 
  40e &  2300.31 &   {\SiII} & 1193.29 &       &  B &   &     \\ 
  41e &  2302.50 &   {\SiII} & 1190.42 &       &  C &   &     \\ 
  42e &  2308.06 &   {\SiII} & 1193.29 &       &  C &   &     \\ 
  43d &  2323.00 &  {\SiIII} & 1206.50 &  33.5 &  A &   &     \\ 
  44d &  2325.77 &  {\SiIII} & 1206.50 &       &  B &   &     \\ 
  45d &  2333.61 &  {\SiIII} & 1206.50 &       &  C &   &     \\ 
  46d &  2340.65 &    {\Lya} & 1215.67 &  13.6 &  A &   &     \\ 
  47d &  2343.45 &    {\Lya} & 1215.67 &       &  B &   &     \\ 
  48d &  2351.35 &    {\Lya} & 1215.67 &       &  C &   &     \\ 
  49d &  2385.23 &     {\NV} & 1238.82 &  97.9 &  A &   & Bl--Galactic {\FeII} \\ 
  50d &  2388.08 &     {\NV} & 1238.82 &       &  B &   &     \\ 
\tablebreak
  51d &  2392.89 &     {\NV} & 1242.80 &       &  A &   &     \\ 
  52d &  2395.75 &     {\NV} & 1242.80 &       &  B &   & Bl--53d \\ 
  53d &  2396.13 &     {\NV} & 1238.82 &       &  C &   & Bl--52d \\ 
  54d &  2403.83 &     {\NV} & 1242.80 &       &  C &   &     \\ 
  55f &  2426.82 &   {\SiII} & 1260.42 &  16.3 &  A &   &     \\ 
  56f &  2429.72 &   {\SiII} & 1260.42 &       &  B &   & Bl--{\Lya}?    \\ 
  57f &  2437.91 &   {\SiII} & 1260.42 &       &  C &   &     \\ 
  58g &  2569.51 &    {\CII} & 1334.53 &  24.4 &  A &   &     \\ 
  59g &  2572.58 &    {\CII} & 1334.53 &       &  B &   & Bl--{\Lya}? \\ 
  60g &  2581.25 &    {\CII} & 1334.53 &       &  C &   & Bl--{\Lya}? \\ 
  61h &  2683.54 &   {\SiIV} & 1393.76 &  45.1 &  A &   & Note\tablenotemark{a}    \\ 
  62h &  2686.74 &   {\SiIV} & 1393.76 &       &  B &   &     \\ 
  63h &  2695.80 &   {\SiIV} & 1393.76 &       &  C &   &     \\ 
  64h &  2700.89 &   {\SiIV} & 1402.77 &       &  A &   &     \\ 
  65h &  2704.12 &   {\SiIV} & 1402.77 &       &  B &   &     \\ 
  66h &  2713.24 &   {\SiIV} & 1402.77 &       &  C &   &     \\ 
  67i &  2980.89 &    {\CIV} & 1548.20 &  64.5 &  A &   &     \\ 
  68i &  2984.46 &    {\CIV} & 1548.20 &       &  B &   & Bl--69i \\ 
  69i &  2985.85 &    {\CIV} & 1550.77 &       &  A &   & Bl--68i \\ 
  70i &  2989.42 &    {\CIV} & 1550.77 &       &  B &   &     \\ 
  71i &  2994.52 &    {\CIV} & 1548.20 &       &  C &   &     \\ 
  72i &  2999.50 &    {\CIV} & 1550.77 &       &  C &   &     \\ 
\enddata
\tablecomments{The letter component to the line number gives the
Figure~3 panel designation. ``C--fit?'' indicates that the continuum
fit is somewhat uncertain.  ``Bl--X'' indicates an identified line
blend.}
\tablenotetext{a}{The clear presence of {\Lysix} from system C at
$1799.2$~{\AA} and of {\SiIV} $\lambda 1393$ from system A at
$2683.5$~{\AA} casts doubt upon the reality of the {\OVI}--selected
system at $z=0.7338$ reported by Burles \& Tytler (1996).}
\label{tab:fosIDs}
\end{deluxetable}

\begin{table}
\tablenum{3}
\rotate[l]{\makebox[0.95\textheight][l]{\vbox{
\begin{center}
\centerline{3. Bracketed Ionization Conditions} 
\vglue 0.1in
\begin{tabular}{lrcrrrcrrrrrrcrrr}\hline\hline
& & & & \multicolumn{6}{c}{Maximal Diffuse} & &
\multicolumn{6}{c}{Minimal Diffuse} \\
& & & & \multicolumn{6}{c}{(Low Ionization Limits)} & &
\multicolumn{6}{c}{(High Ionization Limits)} \\
\cline{5-10} \cline{12-17}
{Sys} &
{Cld} &
$N({\MgII})$ & &
{$U$~} &
{$Z$~} &
{\FeII}/{\MgII} &
$N({\HI})$ &
$N({\CIV})$ &
$S$ [pc] & &
{$U$~} &
{$Z$~} &
{\FeII}/{\MgII} &
$N({\HI})$ &
$N({\CIV})$ &
$S$ [pc] \\
 & & & &
(1) &
(2) &
(3) &
(4) &
(5) &
(6) & &
(1) &
(2) &
(3) &
(4) &
(5) &
(6) \\
\hline
A$^{a}$ & 1 & 11.92 & & $-2.5$ & $-0.2$ & $-3.0$ & $15.0$ & $13.3$ & $90$
                    & & $-2.5$ & $-0.2$ & $-3.0$ & $15.0$ & $13.3$ & $90$ \\
A       & 2 & 12.16 & & $-2.5$ & $-0.2$ & $-3.0$ & $15.2$ & $13.5$ & $150$
                    & & $-2.5$ & $-0.2$ & $-3.0$ & $15.2$ & $13.5$ & $150$ \\
A       & 3 & 12.12 & & $-2.5$ & $-0.2$ & $-3.0$ & $15.2$ & $13.5$ & $140$
                    & & $-2.5$ & $-0.2$ & $-3.0$ & $15.2$ & $13.5$ & $140$ \\
A       & 4 & 12.35 & & $-2.5$ & $-0.2$ & $-3.0$ & $15.4$ & $13.7$ & $230$
                    & & $-2.5$ & $-0.2$ & $-3.0$ & $15.4$ & $13.7$ & $230$ \\
A       & 5 & 12.35 & & $-2.5$ & $-0.2$ & $-3.0$ & $15.1$ & $13.4$ & $110$
                    & & $-2.5$ & $-0.2$ & $-3.0$ & $15.1$ & $13.4$ & $110$ \\
A       & 6 & 11.60 & & $-2.5$ & $-0.2$ & $-3.0$ & $14.7$ & $13.0$ & $40$
                    & & $-2.5$ & $-0.2$ & $-3.0$ & $14.7$ & $13.0$ & $40$ \\ 
\cline{1-3} \cline{5-10} \cline{12-17}
B       & 7  & 11.72 & & $-4.0$ & $-0.6$ & $-0.7$ & $15.7$ & $6.7$  & $1$
                     & & $-2.2$ & $-0.6$ & $-4.0$ & $15.4$ & $13.8$ & $780$ \\
B$^{b}$ & 8  & 13.44 & & $-3.2$ & $ 0.0$ & $-0.7$ & $16.7$ & $13.2$ & $110$
                     & & $-3.2$ & $ 0.0$ & $-0.7$ & $16.7$ & $13.2$ & $110$ \\
B$^{b}$ & 9 & 13.29 & & $-3.4$ & $ 0.0$ & $-1.1$ & $16.5$ & $13.6$ & $210$
                     & & $-3.0$ & $ 0.0$ & $-1.1$ & $16.5$ & $13.6$ & $210$ \\
B       & 10 & 12.60 & & $-3.1$ & $-0.6$ & $-1.6$ & $16.2$ & $12.6$ & $60$
                     & & $-2.3$ & $-0.6$ & $-4.0$ & $16.0$ & $14.6$ & $5200$ \\
B$^{b}$ & 11 & 12.82 & & $-3.2$ & $ 0.0$ & $-0.8$ & $16.1$ & $12.7$ & $30$
                     & & $-3.2$ & $ 0.0$ & $-0.8$ & $16.1$ & $12.7$ & $30$ \\
\cline{1-3} \cline{5-10} \cline{12-17}
C & 12 & 12.05 & & $-3.1$ & $-1.0$ & $-1.1$ & $16.3$ & $12.3$ & $100$
  &              & $-2.0$ & $-0.4$ & $-4.5$ & $16.3$ & $15.3$ & $26000$ \\
\hline
\end{tabular}
\end{center}
Note. --- Systems A and B are $\alpha$--group enhanced by 0.5
dex, whereas System C has solar abundance ratios.
Column 1 is the log of the ionization parameter (see
text).  The number density of hydrogen is given by $\log n_{H} =
-(\log U + 5.2)$.  Column 2 is the metallicity, [$Z/Z_{\odot}$]. Column
3 is the ratio of the {\FeII} and {\MgII} column densities, $\log
N({\FeII}) - \log N({\MgII})$. Columns 4 and 5 are the log of the {\HI}
and {\CIV} column densities in atoms cm$^{-2}$. Column 6 is the linear
depth of the cloud in parsecs. \\
\vglue 0.05in
$^{a}$ Note that the low and high ionization ``limits'' of system A
are presented to be identical.  In fact, the limits are very narrow
due to the contraints provided by the {\SiII} and {\SiIV} profiles 
(see text).\\
$^{b}$ The {\FeII}/{\MgII} ratio of this cloud is fixed by
measurement.  The cloud abundance ratio pattern is assumed solar, 
not $\alpha$--group enhanced (see text).
}}}
\label{tab:cloudpars}
\end{table}

\begin{table}
\tablenum{4}
\rotate[l]{\makebox[0.95\textheight][l]{\vbox{
\begin{center}
\centerline{4. Photoionized {\MgII} Cloud Properties}
\vglue 0.1in
\begin{tabular}{lrrrrrrrrrrrrr}\hline\hline
          &      & \multicolumn{12}{c}{Cloud Number} \\
\cline{3-14}
Property  & IP   & 1 & 2 & 3 & 4 & 5 & 6 & 7 & 8 & 9 & 10 & 11 & 12 \\
          & [eV] \\
\hline
$z_{\rm abs}$ \dotfill & & 0.92501 & 0.92535 & 0.92550 & 0.92595 & 0.92639 & 0.92648 & 
                  0.92720 & 0.92742 & 0.92764 & 0.92780 & 0.92803 & 0.93428 \\ 
$v$ \dotfill           & & $-403.5$  & $-350.7$  & $-327.6$  & $-256.8$  & $-187.9$  & $-174.2$ &
                   $-62.8$  & $-29.0$  & $6.0$  & $30.3$  & $66.0$  & $1039.3$ \\
$\log U$ \dotfill  & & $-2.5$  & $-2.5$  & $-2.5$  & $-2.5$  & $-2.5$  & $-2.5$ &
                   $-2.2$  & $-3.2$  & $-3.0$  & $-2.2$  & $-3.2$  & $-2.6$ \\
$S$ [pc] \dotfill    &   & 90 & 150 & 140 & 230 & 110 & 40 & 780 & 110 & 210 & 5200 & 30 & 1700 \\
$Z$ \dotfill           & & $-0.2$  & $-0.2$  & $-0.2$  & $-0.2$  & $-0.2$  & $-0.2$ &
                   $-0.6$  & $ 0.0$  & $ 0.0$  & $-0.6$  & $ 0.0$  & $-1.0$ \\
$[\alpha/{\rm Fe}]$ \dotfill   & & $+0.5$  & $+0.5$  & $+0.5$  & $+0.5$  & $+0.5$  & $+0.5$ &
                   $+0.5$  & $ 0.0$  & $ 0.0$  & $+0.5$  & $ 0.0$  & $ 0.0$ \\
$N({\HI})$     &  13.6 & 15.0 & 15.2 & 15.2 & 15.4 & 15.1 & 14.7 & 15.4 & 16.7 & 16.5 & 16.2 & 16.1 & 16.4 \\
$N({\rm H})$   & \nodata &  17.7 & 18.0 & 17.9 & 18.1 & 17.8 & 17.4 & 18.4 & 18.5 & 18.6 & 19.3 & 18.0 & 19.5 \\
$N({\MgII})$   &  15.0 & 11.9 & 12.2 & 12.1 & 12.3 & 12.0 & 11.6 & 11.7 & 13.4 & 13.3 & 12.6 & 12.8 & 12.1 \\
$N({\FeII})$   &  16.2 &  8.9 &  9.2 &  9.1 &  9.3 &  9.0 &  8.6 &  7.7 & 12.7 & 12.2 &  8.6 & 12.1 & 9.7 \\
$N({\SiII})$   &  16.3 & 12.1 & 12.4 & 12.3 & 12.6 & 12.2 & 11.8 & 12.0 & 13.9 & 13.8 & 12.8 & 13.3 & 12.3 \\
$N({\CII})$    &  24.4 & 12.7 & 12.9 & 12.9 & 13.1 & 12.8 & 12.4 & 12.7 & 14.6 & 14.4 & 13.5 & 14.0 & 13.4 \\
$N({\NII})$    &  29.6 & 11.9 & 12.2 & 12.1 & 12.3 & 12.0 & 11.6 & 11.8 & 14.0 & 13.8 & 12.7 & 13.4 & 12.7 \\
$N({\FeIII})$  &  30.7 & 11.2 & 11.4 & 11.4 & 11.6 & 11.3 & 10.9 & 10.5 & 13.7 & 13.5 & 11.4 & 13.1 & 12.1 \\
$N({\SiIII})$  &  33.5 & 13.2 & 13.4 & 13.4 & 13.6 & 13.2 & 12.8 & 13.2 & 13.6 & 13.8 & 14.1 & 13.1 & 13.4 \\
$N({\SiIV})$   &  45.1 & 13.0 & 13.3 & 13.2 & 13.4 & 13.1 & 12.7 & 13.1 & 12.9 & 13.3 & 14.0 & 12.4 & 13.0 \\
$N({\NIII})$   &  47.4 & 13.4 & 13.6 & 13.6 & 13.8 & 13.4 & 13.0 & 13.6 & 14.4 & 14.5 & 14.4 & 13.8 & 14.0 \\
$N({\CIII})$   &  47.9 & 13.9 & 14.2 & 14.1 & 14.4 & 14.0 & 13.6 & 14.2 & 15.0 & 15.1 & 15.0 & 14.4 & 14.6 \\
$N({\CIV})$    &  64.5 & 13.3 & 13.5 & 13.5 & 13.7 & 13.4 & 13.0 & 13.8 & 13.2 & 13.6 & 14.6 & 12.7 & 13.8 \\
$N({\SVI})$    &  88.0 & 11.4 & 11.7 & 11.6 & 11.8 & 11.5 & 11.1 & 12.2 & 10.5 & 11.1 & 13.0 & 10.1 & 11.9 \\
$N({\NV})$     &  97.9 & 11.7 & 12.0 & 11.9 & 12.1 & 11.8 & 11.4 & 12.5 & 11.3 & 11.8 & 13.3 & 10.8 & 12.1 \\
$N({\OVI})$    & 138.1 & 11.4 & 11.6 & 11.6 & 11.8 & 11.5 & 11.1 & 12.4 &  9.6 & 10.6 & 13.2 &  9.3 & 11.7 \\
\hline
\end{tabular}
\end{center}
Note. --- The redshifts, velocities, and {\MgII} column
densities are measured from the VP decomposition, as are the column
densities of {\FeII} in clouds 8, 9, and 11.  All velocities are
computed with respect to $z = 0.92760$.  All other quantities are
based upon photoionization modeling of the constraint transitions (those with IP
less than that of {\CIV}).  The $\alpha$--group enhancement and
metallicity are inversely proportional for a fixed $N({\HI})$ and
$\log U$.  Without $\alpha$--group enhancement, many clouds would require
supersolar [Fe/H].
}}}
\label{tab:PICs}
\end{table}

\begin{deluxetable}{lcccccccccl}
\tablenum{5}
\tablecolumns{9}
\tablewidth{0pc}
\tablecaption{Diffuse Component Cloud Properties}
\tablehead
{
                   &  & \multicolumn{4}{c}{System A} & 
 & \colhead{System B} & & \colhead{System C} \\ 
\cline{3-6} \cline{8-8} \cline{10-10}
                   &  & \colhead{Single Phase} & & \multicolumn{2}{c}{Two Phase}
               &  & \colhead{Single Phase} & & \colhead{Single Phase} \\ \cline{5-6}
\colhead{Property} &  & \colhead{Photo} & & \colhead{Photo} &
\colhead{Coll} &  & \colhead{Photo} & & \colhead{Photo}
}
\startdata
$z_{\rm abs}$          \dotfill & & 0.92572  & & 0.92572 & 0.92572 & & 0.92768  & & 0.93428  & \\
$Z$                    \dotfill & &    0     & &     0   &     0   & &    0     & & $-0.4$ &  \\
$[{\rm C}/{\rm O}]$    \dotfill & &    0     & &     0   &     0   & &    0     & &   0  &  \\
$b$                    \dotfill & &   70     & &    70   &    70   & &   70     & & 40 & km~s$^{-1}$ \\
$N({\SiIV})$           \dotfill & & 11.3     & &  12.1   &  12.0   & & 12.8     & & 10.3 & cm$^{-2}$ \\
$N({\NIII})$           \dotfill & & 13.3     & &  13.7   &  12.5   & & 14.3     & & 12.8 & cm$^{-2}$ \\
$N({\CIII})$           \dotfill & & 13.8     & &  14.2   &  11.8   & & 14.8     & & 13.4 & cm$^{-2}$ \\
$N({\CIV})$            \dotfill & & 14.5     & &  14.5   &  13.5   & & 15.2     & & 13.9 & cm$^{-2}$ \\
$N({\SVI})$            \dotfill & & 12.7     & &  13.1   &  13.4   & & 13.7     & & 12.1 & cm$^{-2}$\\
$N({\NV})$             \dotfill & & 14.2     & &  13.8   &  14.2   & & 14.5     & & 13.6 & cm$^{-2}$ \\
$N({\OVI})$            \dotfill & & 15.0     & &  14.3   &  15.0   & & 14.9     & & 14.5 & cm$^{-2}$\\
$\log U$               \dotfill & & $-1.2$   & &  $-1.6$ & \nodata & & $-1.6$   & & $-1.3$ & \\
$N({\HI})$             \dotfill & & 14.6     & &  14.8   &  13.5   & & 15.5     & & 14.5 & cm$^{-2}$ \\
$N({\rm H})$           \dotfill & & 18.7     & &  18.5   &  19.2   & & 19.1     & & 18.6 & cm$^{-2}$ \\
$\log n_{H}$\tablenotemark{a}  \dotfill & & $-4.0$   & &  $-3.6$ & \nodata & & $-3.6$   & & $-3.9$ & cm$^{-3}$ 
\\
$S$\tablenotemark{a}          \dotfill & & 16.5     & &  3.9    & $0.01/n_{H}$ & & 16.0     & & 11.6 & kpc \\
$N({\CIV})/N({\NV})$   \dotfill & & 1.9      & &  5.0    & 0.2     & & 5.0           & & 2.0 & \\
$N({\CIV})/N({\OVI})$  \dotfill & & 0.3      & &  1.6    & 0.03    & & 1.6      & & 0.3  & \\
$N({\CIV})/N({\SiIV})$ \dotfill & & 1320     & &  250    & 30      & &220       & & 4000 & \\
\enddata
\tablenotetext{a}{For a fiducial density range of $-3 \leq \log n_{H}
[{\cc}] \leq -4$ for the collisionally ionized phase of system A, the
inferred size is $10 \leq S \leq 100$~kpc.}
\label{tab:DICs}
\end{deluxetable}


\end{document}